\documentclass[hyper,a4paper]{JHEP3}

\usepackage{graphicx}
\usepackage{graphics}
\usepackage[mathscr]{eucal}
\usepackage{amssymb}
\usepackage{amsmath}
\usepackage{xspace}
\usepackage{listings}
\usepackage{ulem}
\usepackage{slashed}
\usepackage{verbatim}
\usepackage{axodraw4j}
\usepackage{caption}
\usepackage{subcaption}
\usepackage{multirow}

\def\mpt{{\slash\!\!\!\!\!\:P}_T}

\def\mptv{{\slash\!\!\!\!\!\:\vec{P}}_T}

\title{Discovery potential for $T' \to tZ$ in the trilepton channel at the LHC}

\author{
Lorenzo Basso$^{a\ast}$ and Jeremy Andrea$^{b}$\\
Universit\`e de Strasbourg, IPHC, 23 rue du Loess 67037 Strasbourg,
      France \\
Institut Pluridisciplinaire Hubert Curien/D\'epartement Recherches 
Subatomiques,
CNRS-IN2P3, 23 Rue du Loess, F-67037 Strasbourg, France

$^a$Email: \email{lorenzo.basso@iphc.cnrs.fr} \\
$^b$Email: \email{jeremy.andrea@iphc.cnrs.fr} \\
$^\ast$Corresponding author\\
}
\abstract{The LHC discovery potential of heavy top partners decaying into a top quark and a $Z$ boson is studied in the trilepton channel at $\sqrt{s}=13$ TeV in the single production mode. The clean multilepton final state allows to strongly reduce the background contaminations and to reconstruct the $T'$ mass. We show that a simple cut-and-count analysis probes the parameter space of a simplified model as efficiently as a dedicated multivariate analysis. The trilepton signature finally turns out to be as sensitive in the low $T'$ mass region as the complementary channel with a fully hadronic top quark, and more sensitive in the large mass domain. The reinterpretation in terms of the top-$Z$-quark anomalous coupling is shown.}


\begin{document}

\section{Introduction}
In 2012 the Run-I of the LHC at the center of mass energies of $7$ and $8$ TeV has finally discovered the long sought-after Higgs boson~\cite{Chatrchyan:2012ufa,Aad:2012tfa}. In Spring 2015 the LHC will start again to produce $pp$ collisions in the so-called Run-II, at the increased energy of $13$ TeV. It is expected to accumulate $100-150$ inverse femtobarn (fb$^{-1}$) of data in the first two years and 
up to $300$ fb$^{-1}$ in the following ones.
The primary scope of the Run-II is to further understand the newly discovered Higgs boson and to eventually make new discoveries. 
In all generality, it is very common in beyond the standard model theories that new heavy fermions arise to stabilise the Higgs boson mass and to protect it from dangerous quadratic divergences. In many cases, these new fermions are heavy partners of the third generation quarks with vector-like couplings. They are commonly predicted by many new physics scenarios, including Extra Dimensions, Little Higgs Models, and Composite Higgs Models~\cite{Contino:2006qr,Contino:2006nn,Contino:2008hi,Cabarcas:2008ys,Anastasiou:2009rv,Vignaroli:2012nf,Matsedonskyi:2012ym,Ellis:2014dza}. The observation of new heavy quarks thus plays an important role in the investigation of the Higgs sector. The common feature of these heavy quarks is to decay into a standard model quark and a $W^\pm$ boson, a $Z$ bosons, or a Higgs boson.\footnote{Other possibilities  like 3-body decays are also possible, but are not covered here.} The relative branching ratios are purely determined by the weak quantum numbers of the multiplet the new quark belongs to~\cite{delAguila:2000rc}. Here we will focus on the case of a singlet heavy quark: the {\it{top partner}} or $T'$. Experimental collaborations are now considering the interplay between the various decay channels in their searches. Recent limits from ATLAS~\cite{Aad:2014efa} and CMS~\cite{Chatrchyan:2013uxa} lie within 690-780 GeV, depending on the considered final state.

These searches however do not typically consider intergenerational mixing. 
Although vector-like quarks are usually assumed to only mix with the third generation following hierarchy or naturalness arguments~\cite{Berger:2012ec,DeSimone:2012fs,Aguilar-Saavedra:2013wba,Aguilar-Saavedra:2013qpa,Han:2014qia}, the top partners can mix in a sizable way with lighter quarks while remaining compatible with the current experimental constraints~\cite{Cacciapaglia:2011fx,Buchkremer:2013bha}. This possibility has been recently pursued and must be considered with attention. The top partners interactions with the electroweak and Higgs bosons are generically allowed through arbitrary Yukawa couplings, implying that the branching ratios into light quarks can be possibly competitive with the top-quark one. Here we consider only the case of mixing to the first generation of quarks.

Beside opening up the decay channel into a standard model boson plus a light quark, the mixing with the first generation also enhances the single production, especially due to the presence of valence quarks in the initial state. Even without mixing, the single production cross sections at the upcoming LHC energies become competitive with the pair production ones. Furthermore, the $T' \to tZ$ final state is only recently being experimentally investigated, both in the dilepton and in the trilepton channels by ATLAS~\cite{Aad:2014efa}. Following the investigation first pursued at the 2013 Les Houches workshop~\cite{Brooijmans:2014eja}, in this paper we study the LHC discovery potential of the $T' \to tZ$ channel in the trilepton decay mode in single production at $\sqrt{s}=13$ TeV, for a singlet $T'$ quark mixing with the first generation.  To capture all the essential features of the new heavy top quark while remaining as model independent as possible, the study here presented is performed in the framework of simplified models. We will employ the dedicated model for the heavy singlet top partner as presented in ref.~\cite{Buchkremer:2013bha}, comprising only $3$ independent couplings.

The paper is structured as follows. In section~\ref{sect:model} the simplified model under consideration is described. In addition, we propose a convenient way to explore its parameter space. In section~\ref{sect:pheno} we describe our analysis, comparing the discovery reach obtained in a simple cut-and-count approach to that obtained in a dedicated multivariate analysis. Results for $\mathcal{L}=100$ fb$^{-1}$ are collected in section~\ref{sect:results}. The main result of this paper is that the simple cut-and-count analysis probes the $T'$ parameter space as efficiently as the more sophisticated multivariate analysis. Also, a comparison to the complementary dilepton channel, recently presented in~\cite{Reuter:2014iya} for a $T'$ coupling only to the third generation, is pursued. 
The reinterpretation of our results to the case of the top-quark flavour-changing neutral coupling to a light quark and a $Z$ boson, which shares the same final state as the $T'$ discussed here, is overviewed in section~\ref{sect:FCNC}.
Finally, we conclude in section~\ref{sect:conclusions}.

\section{Simplified model}\label{sect:model}
A simple Lagrangian that parametrises the $T'$ couplings to quarks and electroweak boson is (showing only the couplings relevant for our analysis)~\cite{Buchkremer:2013bha}
\begin{eqnarray}\label{Tp_lag}
\mathcal{L}_{\rm T'} &=& g^\ast \left\{ \sqrt{\frac{R_L}{1+R_L}} \frac{g}{\sqrt{2}} [ \overline{T'}_{L/R} W^+_\mu \gamma^\mu d_{L/R} ] + \sqrt{\frac{1}{1+R_L}} \frac{g}{\sqrt{2}} [ \overline{T'}_{L/R} W^+_\mu \gamma^\mu b_{L/R} ] + \right.   \label{eq:lagrangian} \\ \nonumber
&& +\left.  \sqrt{\frac{R_L}{1+R_L}} \frac{g}{2 \cos \theta_W} [ \overline{T'}_{L/R} Z_\mu \gamma^\mu u_{L/R} ] + \sqrt{\frac{1}{1+R_L}} \frac{g}{2 \cos \theta_W} [ \overline{T'}_{L/R} Z_\mu \gamma^\mu t_{L/R} ]  \right\} + h.c.\, , 
\end{eqnarray}
where the subscripts $L$ and $R$ label the chiralities of the fermions. 
Only $3$ parameters are sufficient to fully describe the interactions that are relevant for our investigation. Besides $M_{T'}$, the vector-like mass of the top partner, there are the $2$ couplings appearing in eq.~(\ref{Tp_lag})

\begin{itemize}
\item[-] $g^\ast$, the coupling strength to SM quarks in units of standard couplings, which is only relevant in single production. The cross sections for the latter scale with the coupling squared;
\item[-] $R_L$, the generation mixing coupling, which describes the rate of decays to first generation quarks with respect to the third generation, so that $R_L = 0$ corresponds to coupling to top and bottom quarks only, while the limit $R_L = \infty$ represents coupling to first generation of quarks only.
\end{itemize}

For some possible reinterpretation of this effective Lagrangian in terms of complete models, see refs.~\cite{Ellis:2014dza,Reuter:2014iya}.

\subsection{Cross section parameterisation}
\begin{figure}[!t]
\begin{subfigure}{0.48\textwidth}
\centering
\scalebox{0.7}{
  \begin{picture}(228,188) (175,-104)
    \SetWidth{1.0}
    \SetColor{Black}
    \Line[arrow,arrowpos=0.5,arrowlength=12.5,arrowwidth=5,arrowinset=0.2](176,60)(256,44)
    \Line[arrow,arrowpos=0.5,arrowlength=12.5,arrowwidth=5,arrowinset=0.2](256,44)(336,60)
    \Photon(256,44)(256,-68){7.5}{6}
    \Line[arrow,arrowpos=0.5,arrowlength=12.5,arrowwidth=5,arrowinset=0.2](336,-68)(400,-100)
    \Photon(336,-68)(400,-20){7.5}{4}
    \Line[arrow,arrowpos=0.5,arrowlength=12.5,arrowwidth=5,arrowinset=0.2](176,-84)(256,-68)
    \SetWidth{1.5}
    \Line[arrow,arrowpos=0.5,arrowlength=12.5,arrowwidth=5,arrowinset=0.2,double,sep=3.5](256,-68)(336,-68)
    \Text(210,71)[r]{\LARGE $q$}
    \Text(310,71)[r]{\LARGE $q\slash q'$}
    \Text(220,-100)[r]{\LARGE $u\slash d$}
    \Text(320,0)[r]{\LARGE $Z\slash W^\pm$}
    \Text(310,-48)[r]{\LARGE $T'$}
    \Text(370,-20)[r]{\LARGE $Z$}
    \Text(370,-100)[r]{\LARGE $t$}
  \end{picture}}
\caption{$\mathcal{A}_1$}
\end{subfigure}
\begin{subfigure}{0.48\textwidth}
\centering
\scalebox{0.7}{
  \begin{picture}(228,188) (175,-104)
    \SetWidth{1.0}
    \SetColor{Black}
    \Line[arrow,arrowpos=0.5,arrowlength=12.5,arrowwidth=5,arrowinset=0.2](176,60)(256,44)
    \Line[arrow,arrowpos=0.5,arrowlength=12.5,arrowwidth=5,arrowinset=0.2](256,44)(336,60)
    \Photon(256,44)(256,-68){7.5}{6}
    \Line[arrow,arrowpos=0.5,arrowlength=12.5,arrowwidth=5,arrowinset=0.2](336,-68)(400,-100)
    \Photon(336,-68)(400,-20){7.5}{4}
    \Line[arrow,arrowpos=0.5,arrowlength=12.5,arrowwidth=5,arrowinset=0.2](176,-84)(256,-68)
    \SetWidth{1.5}
    \Line[arrow,arrowpos=0.5,arrowlength=12.5,arrowwidth=5,arrowinset=0.2,double,sep=3.5](256,-68)(336,-68)
    \Text(210,71)[r]{\LARGE $q$}
    \Text(310,71)[r]{\LARGE $q'$}
    \Text(220,-100)[r]{\LARGE $b$}
    \Text(300,0)[r]{\LARGE $W^\pm$}
    \Text(310,-48)[r]{\LARGE $T'$}
    \Text(370,-20)[r]{\LARGE $Z$}
    \Text(370,-100)[r]{\LARGE $t$}
  \end{picture}}
\caption{$\mathcal{A}_3$}
\end{subfigure}
\caption{Feynman diagrams for the process $pp\to T'j\to tZj$ via couplings of the $T'$ to (a) first generation quarks and (b) third generation quarks.\label{fig:diagrams}}
\end{figure}
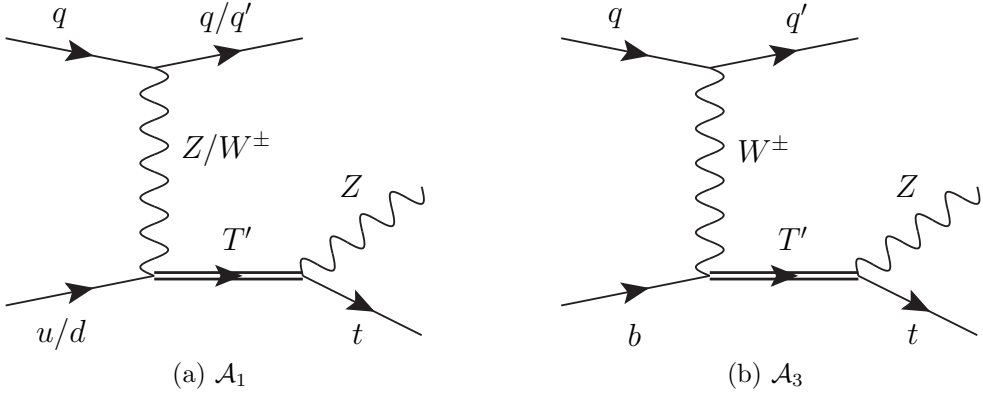

In this paper we study the LHC discovery potential of the $T'$ in single production mode, in association with a light jet. Then, the $T'$ decays into a top and a $Z$ boson. The overall signature reads: $pp\to T'j\to tZj$. This process is given by the set of Feynman diagrams displayed in figure\ref{fig:diagrams}.

The are two sets of diagrams, i.e. where the $T'$ is produced due to the interaction with light quarks ($\mathcal{A}_1$) or due to the interaction with the $b$ quark ($\mathcal{A}_3$). From the Lagrangian in eq.~(\ref{eq:lagrangian}), these 2 sets of diagrams give production cross sections that scale differently with $R_L$, the mixing coupling. Further, the decay into a top quark and a Z boson scales with $R_L$ too. We parametrise the production cross section and branching ratio BR($T' \to tZ$) as follows:
\begin{eqnarray}\label{eq:rescaling_xs}
\sigma_{pp\to T'}(M_{T'},\,R_L) &=& \mathcal{A}_1(M_{T'})\, \frac{R_L}{1+R_L} + \mathcal{A}_3(M_{T'})\, \frac{1}{1+R_L}\, ,\\ \label{eq:rescaling_br}
BR_{T'\to tZ}(M_{T'},\,R_L) &=& \mathcal{B}(M_{T'})\, \frac{1}{1+R_L}\, ,
\end{eqnarray}
where $\mathcal{A}_i(M_{T'})$ $(i=1,3)$ and $\mathcal{B}(M_{T'})$ are numerical coefficients. In details, $\mathcal{A}_i$ represents the production cross section for the heavy quark due to interactions to partons belonging to the $i^{th}$ generation, while $\mathcal{B}$ is the $T'$ branching ratio for its decay into a top quark and a Z boson.\footnote{Eq.~(\ref{eq:rescaling_br}) holds when all the $T'$ decay products are much lighter than its mass, which is the case under consideration.} They have been evaluated at the LHC at $\sqrt{s}=13$ TeV for the {\tt CTEQ6L1} PDF~\cite{Pumplin:2002vw} and collected in table~\ref{tab:xs_coeff}, for the first and third generations only.

\TABLE{
\centering
 \begin{tabular}{|c|c|c|c|}
 \hline
  $M_{T'}$ (GeV) & $\mathcal{A}_1(M_{T'})$ (pb) & $\mathcal{A}_3(M_{T'})$ (pb) & $\mathcal{B}(M_{T'})$ (\%)  \\
 \hline
 800  & 1.2614 & 0.07242 & 22.4 \\
 1000 & 0.7752 & 0.03518 & 23.5 \\
 1200 & 0.5001 & 0.01826 & 24.0 \\
 1400 & 0.3331 & 0.00994 & 24.2 \\
 1600 & 0.2265 & 0.00561 & 24.4 \\
 \hline
 \end{tabular}
 \caption{Numerical coefficients for $1^{st}$ and $3^{rd}$ production cross sections at the LHC at $\sqrt{s}=13$ TeV and $T'$ branching ratios to $tZ$.}
 \label{tab:xs_coeff}
}

These formulas will allow us to draw the LHC discovery power curves as a function of $R_L$ in section~\ref{sect:results} in a simple way. The product of eq.~(\ref{eq:rescaling_xs}) and eq.~(\ref{eq:rescaling_br}) gives the cross section for $pp\to T'j \to tZj$ as a function of $R_L$ in the narrow-width approximation. However, this also depends on the choice of the PDF, on the top mass, the EW couplings, etc. By considering the ratio of the cross sections evaluated at different values of $R_L$ (and all other input fixed) one can factorise the impact of the former. Such ratio can then be used to rescale a given cross section, evaluated at the user's choice of the input parameters, as a function of $R_L$. In our case, we computed the cross section $\sigma (M_{T'},\,0.5)$ of the various benchmark points for $R_L=0.5$, hence their cross section for an arbitrary value of $R_L$ is given by
\begin{equation}\label{eq:rescaling}
\sigma (M_{T'},\,R_L)=\sigma (M_{T'},\,0.5)\, \frac{\sigma_{pp\to T'}(M_{T'},\,R_L)\, BR_{T'\to tZ}(M_{T'},\,R_L)}{\sigma_{pp\to T'}(M_{T'},\,0.5)\,BR_{T'\to tZ}(M_{T'},\,0.5)}\, .
\end{equation}
Furthermore, this rescaling can be applied anywhere in the cutflow. It is sufficient to replace $\sigma (M_{T'},\,0.5)$ with the cross section after any cut, to obtain the corresponding cross section at a different value of $R_L$ after the same cut.

\section{Analysis strategy}\label{sect:pheno}
All samples employed in this study have been generated in {\tt MadGraph5\_aMC@NLO~v2.1.2}~\cite{Alwall:2014hca} with the {\tt CTEQ6L1} PDF~\cite{Pumplin:2002vw}. Events have subsequently been hadronised/parton showered in {\tt PYTHIA~6}~\cite{Sjostrand:2006za} with tune Z2~\cite{Field:2011iq}. Detector simulation is performed with a customised version of {\tt Delphes~3}~\cite{deFavereau:2013fsa} to emulate the CMS detector. Jets have been reconstructed with {\tt FastJet}~\cite{Cacciari:2011ma} employing the anti-$k_t$ algorithm~\cite{Cacciari:2008gp} with parameter $R=0.5$.

The signal (S) is generated at leading order from the model implemented in {\tt FeynRules}~\cite{Alloul:2013bka,FeynRules_vlq}.
We generate $5$ benchmark points varying the $T'$ mass in steps of $200$ GeV in the range
\begin{equation}
M_{T'} \in \left[ 800;1600\right] \mbox{ GeV,}
\end{equation}
with $g^\ast=0.1$ and $R_L = 0.5$. We do not apply here any $k$-factor to the signal. Contrary to the backgrounds, tau leptons have not been included in the generation of the signal samples.

Backgrounds (B) that can give 3 leptons in the final state which are considered in this analysis are: $t\overline{t}$ and $Z/W+jets$ with non-prompt leptons, and $t\overline{t}W$, $t\overline{t}Z$, $tZj$ and $VZ$ ($V=W,Z/\gamma$) with only genuinely prompt leptons. We generated leading order samples with up to 2 merged jets normalised to the (N)NLO cross section where available, taken from~\cite{Alwall:2014hca,Czakon:2013goa}.
For both signal and background, a suitable number of unweighted event is generated. We checked that the statistical uncertainties at any point of our analyses are below $1\%$, and for this reason they will not be quoted.

We do not simulate multijet backgrounds, which can be reliably estimated only from data. Further, we do not consider jets faking electrons, since this is a feature which is not supported in {\tt Delphes}. It is however seen to be negligible in multileptons analyses (see for instance, ref.~\cite{Khachatryan:2014ewa}).

The analysis is carried out in {\tt MadAnalysis~5}~\cite{Conte:2012fm,Conte:2014zja}.
Leptons ($\ell = e,\, \muÂ$) and jets are identified if passing the following criteria:
\begin{eqnarray}
p_T(\ell) > 20 {\mbox{ GeV,}} &\qquad & |\eta(e/\mu)| < 2.5/2.4\, ,\\
p_T(j) > 40 {\mbox{ GeV,}} &\qquad & |\eta(j)| < 5\, , \\
\Delta R(\ell, j) > 0.4 &\qquad & 
\end{eqnarray}

External routines for $b$-tagging and for lepton isolation have been implemented. Regarding the former, here we adopted the medium working point~\cite{Chatrchyan:2012jua}, which has an average $b$-tagging rate of $70\%$ and a light mistag rate of $1\%$. To apply the $b$-tagging, we considered jets within the tracker only, i.e. with $|\eta(j)| < 2.4$. This means that the $b$-tagging probability for jets with larger pseudorapidities is vanishing. For the latter, the combined tracker-calorimetric isolation is used to identify isolated leptons. The relative isolation $I_{rel}$ is defined as the sum of the $p_T$ and calorimetric deposits of all tracks within a cone of radius $\Delta R=0.3$, divided by the $p_T$ of the lepton. The latter is isolated if $I_{rel}\leq 0.10$. This choice has been taken as a compromise to strongly reduce backgrounds with non-prompt leptons without suppressing the signal, where the two leptons coming from the $Z$ boson get closer and closer as the top partner mass increases.

After the object reconstruction and selection, we apply some general preselections as follows: we require at least 1 jet and no more than 3, of which exactly one is b-tagged, and exactly 3 leptons (electrons or muons). The requirement of less than 3 jets removes the $T'$ pair production isolating the single production channel.

Efficiencies and event yields are evaluated for $\mathcal{L}=100$ fb$^{-1}$ and are collected in table~\ref{tab:presel_eff}. 
The requirement of $3$ isolated leptons strongly reduces the $t\overline{t}+X$ backgrounds, especially the $t\overline{t}+ jets$ one. The diboson component is instead suppressed by the $b$-tagging. Regarding the signal, the requirement of $3$ isolated leptons is less efficient as the $T'$ mass increases. This is because the $2$ leptons stemming from the $Z$ boson gets closer to each other as the $T'$ gets heavier, due to the larger boost of the $Z$ boson in the $T'\to t\,Z$ decay.

\TABLE{
\centering
 \begin{tabular}{|c|c|c|c|c|}
 \hline
 Background  & no cuts & $1\leq n_j\leq 3$  & $n_\ell \equiv 3$ & $n_b \equiv 1$  \\
 \hline
 $t\overline{t}(+X)$  & $7.5 \, 10^{6}$~($100\%$) & $6.1 \, 10^{6}$~($81.2\%$) & 514.9~($0.09\%$) & 243.8~($47.3\%$)\\
 $tZj$     & 3521~($100\%$) & 2953~($83.9\%$) & 290.6~($9.8\%$) & 170.0~($58.5\%$)\\
 $WZ$      & $1.4 \, 10^{5}$~($100\%$) & $5.7 \, 10^{4}$~($41.9\%$) & 3883~($6.9\%$) & 164.3~($4.2\%$)\\ \hline
 Total & $7.6 \, 10^{6}$~($100\%$) & $6.1 \, 10^{6}$~($80.5\%$) & 4689~($0.08\%$) & 578.0~($12.3\%$)\\ \hline\hline

 $M_{T'}$ (GeV) & no cuts & $1\leq n_j\leq 3$  & $n_\ell \equiv 3$ & $n_b \equiv 1$  \\ \hline
 800    & 119.7~($100\%$) & 105.0~($87.8\%$) & 39.3~($37.4\%$) & 25.5~($64.8\%$)\\
 1000   & 77.1~($100\%$) & 67.8~($87.9\%$) & 26.0~($38.4\%$) & 16.4~($63.2\%$)\\
 1200   & 52.0~($100\%$) & 45.3~($87.2\%$) & 16.1~($35.6\%$) & 10.1~($62.4\%$)\\
 1400   & 35.3~($100\%$) & 30.5~($86.6\%$) & 8.0~($26.1\%$)  & 4.8~($60.1\%$)\\
 1600   & 24.5~($100\%$) & 21.1~($86.0\%$) & 3.8~($18.0\%$)  & 2.2~($58.3\%$)\\
 \hline
 \end{tabular}
 \caption{Events surviving the preselections and relative efficiencies (with respect to the previous item). Signal computed for $g^\ast = 0.1$ and $R_L = 0.5$.}
 \label{tab:presel_eff}
}

Finally, the pair of same-flavour and opposite-sign leptons closest to the $Z$ boson mass is chosen, and a cut around their invariant mass distribution is performed,
\begin{equation}\label{eq:Z-cut}
|M (\ell^+ \ell^-) - M_Z| < 15 \mbox{ GeV}\, .
\end{equation}
The lepton from the top decay is therefore identified as the remaining one in our trilepton channel and labelled $\ell_W$.

We describe in the following the 2 analyses we performed, that differentiate from this point on. The first one is a traditional cut-and-count strategy, where subsequent cuts are applied to the most important kinematic variables to maximise the signal-over-background ratio. The second one is a multivariate analysis (MVA), where several discriminating observables are used at once to distinguish the signal from the background, cutting at the end only on its output.


\subsection{Cut-and-count}\label{sect:cc}
The first strategy to study the LHC discovery potential illustrated here is the cut-and-count one (C\&C).
As for the previous reconstruction of the $Z$ boson from its decay products, one could also reconstruct both the top quark and the $W$ boson stemming from its decay. The presence of only one source of missing energy (one neutrino) allows to reconstruct the four momentum of the latter by imposing suitable kinematical constraints. Hence one can reconstruct the $W$ boson and the top quark as resonances in the invariant mass as well as in the transverse mass distributions of the decay products. We chose to use the latter, in the following formulation~\cite{Barger:1987du}:
\begin{equation}
m^2_T = \left( \sqrt{M^2(vis)+P^2_T(vis)}+\left| \mpt \right| \right) ^2
	- \left( \vec{P}_T(vis) + \mptv\right) ^2\, ,
\end{equation}
because of the sharper peaks as compared to those in the invariant mass distributions. Instead to draw suitable window mass cuts, we decided to only apply loose selections, which maximise their efficiencies while retaining most of the signal.
The cuts we applied are as follows:
\begin{eqnarray}\label{eq:W-cut}
10 < &M_T (\ell_W) / \mbox{GeV}& < 150 \, , \\ \label{eq:top-cut}
 0 < &M_T (\ell_W\, b) / \mbox{GeV}& < 220 \, .
\end{eqnarray}
In particular, the lower cut in eq.~(\ref{eq:W-cut}) is inspired by experimental analyses to suppress the multijet background, which we did not simulate.

Surviving events and relative efficiencies are collected in table~\ref{tab:cc_eff}. The numerical values of the cuts appearing in eqs.~(\ref{eq:Z-cut})--(\ref{eq:top-cut}) have been chosen to maximise the signal-over-background ratio while keeping at least $90\%$ of the signal.

\TABLE{
\centering
 \begin{tabular}{|c|c|c|c|c|}
 \hline
 Background  & $n_b \equiv 1$ & cut~(\ref{eq:Z-cut}) & cut~(\ref{eq:W-cut}) & cut~(\ref{eq:top-cut})  \\
 \hline
 $t\overline{t}(+X)$  & 243.8~($47.3\%$) & 154.8~($63.5\%$) & 135.1~($87.3\%$) & 83.0~($61.5\%$)\\
 $tZj$     & 170.0~($58.5\%$) & 155.6~($67.2\%$) & 148.7~($95.6\%$) & 139.8~($63.7\%$)\\
 $WZ$      & 164.3~($4.2\%$)  & 146.9~($89.4\%$) & 138.2~($94.1\%$) & 71.5~($51.7\%$)\\ \hline
 Total     & 578.0~($12.3\%$) & 457.2~($79.1\%$) & 422.0~($92.3\%$) & 294.3~($69.8\%$)\\ \hline\hline

 $M_{T'}$ (GeV) & $n_b \equiv 1$ & cut~(\ref{eq:Z-cut}) & cut~(\ref{eq:W-cut}) & cut~(\ref{eq:top-cut})  \\
 \hline
 800    & 25.5~($64.8\%$) & 23.8~($93.6\%$) & 22.2~($93.2\%$) & 20.8~($93.6\%$)\\
 1000   & 16.4~($63.2\%$) & 15.4~($93.8\%$) & 14.3~($92.4\%$) & 13.4~($94.0\%$)\\
 1200   & 10.1~($62.4\%$) &  9.5~($94.2\%$) &  8.7~($92.3\%$) &  8.1~($92.3\%$)\\
 1400   &  4.8~($60.1\%$) &  4.5~($93.5\%$) &  4.1~($92.1\%$) &  3.8~($91.3\%$)\\
 1600   &  2.2~($58.3\%$) &  2.1~($93.3\%$) &  1.9~($92.2\%$) &  1.7~($90.0\%$)\\
 \hline
 \end{tabular}
 \caption{Object selection surviving events (and efficiencies with respect to the previous item). Signal computed for $g^\ast = 0.1$ and $R_L = 0.5$.}
 \label{tab:cc_eff}
}

Contrary to ref.~\cite{Chatrchyan:2013uxa}, we do not require a forward jet to not suppress any further the signal, despite it being a distinctive feature of our signature. This is also not necessary: the signal is already clearly visible above the background in the distribution of the transverse mass of the $T'$ decay products (the $3$ charged leptons and the $b$-jet), as can been seen in figure~\ref{fig:MtZ} for the various signal benchmark points.

\FIGURE{
\centering
\includegraphics[width=\linewidth]{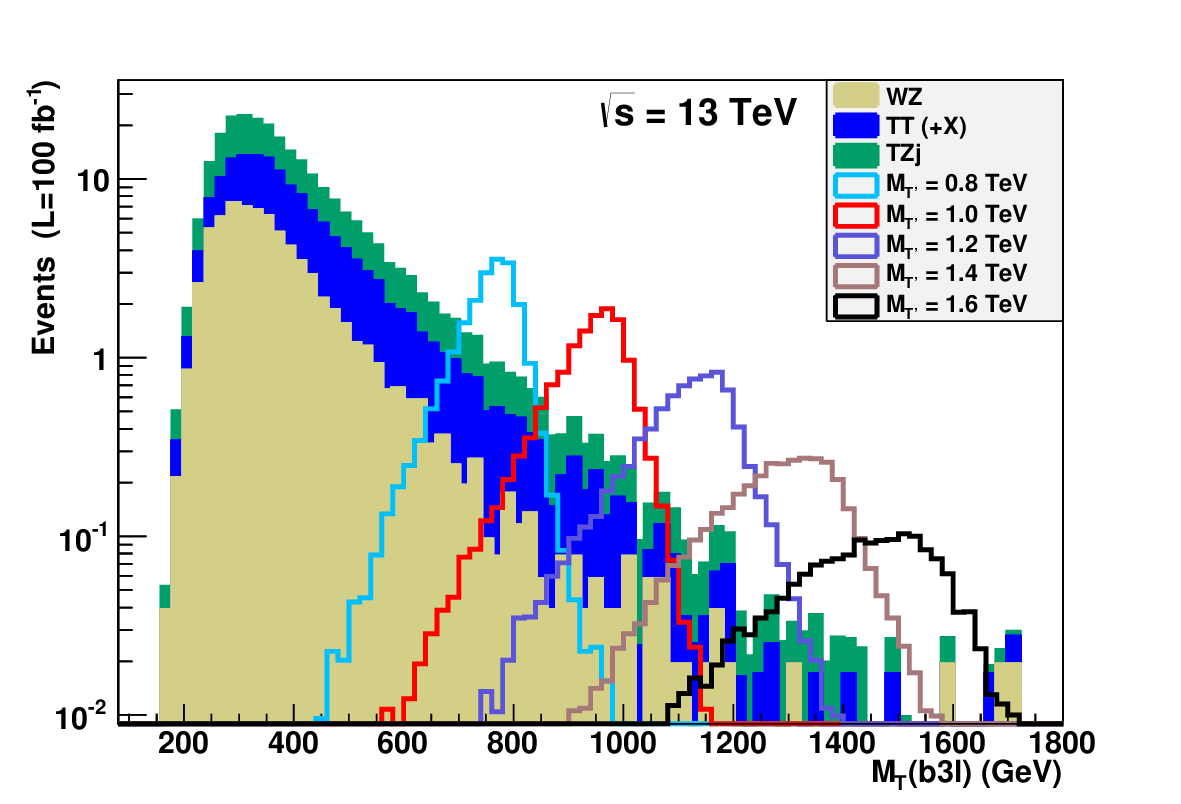} 
\caption{Transverse mass distribution for the $T'$ decay products: the $3$ charged leptons and the $b$-jet.}
\label{fig:MtZ}
}

We select a window around the peak of each benchmark point to get the best signal-over-background significance and we collect the final numbers in table~\ref{tab:significances}.


\subsection{Multivariate analysis}\label{sect:MVA}
In section~\ref{sect:cc} we showed that suitable cuts on the most straightforward distributions were sufficient to isolate the signal from the background. We made use of the signal topology, which has a $Z$ boson decaying leptonically, and a top-quark decaying into a $b$-jet and into a leptonic $W$-boson. The presence of the intermediate $T'$ was then seen as a peak in the transverse mass of all its visible decay products. One could wonder if this was the best strategy, i.e. cutting on those variables with the values we chose.
There are in fact many additional variables that one could analyse to distinguish the signal from the background. However, cutting on any of these variables will unavoidably reduce also the signal. To overcome this,
several variables can be combined using a multivariate analysis (MVA)
to obtain the best signal/background discrimination~\cite{Hocker:2007ht}. We identified some discriminating variables in table~\ref{tab:MVA_var}, ranked according to their discriminating power when a boosted decision tree (BDT) is employed. Here, $\Delta\varphi$ is the difference of the azimuthal angles between 2 objects, $\Delta\eta$ is the difference of their pseudorapidities, and $\Delta R = \sqrt{(\Delta\varphi)^2 + (\Delta\eta)^2}$.

\TABLE{
\centering
 \begin{tabular}{|c|c|c|c|}
 \hline
 Variable  & Importance &  Variable  & Importance   \\
 \hline
$M_T(b\,3\ell)$ & $2.60\, 10^{-1}$ & $\Delta R (b,\, \ell_W)$ & $9.77\, 10^{-2}$ \\
$p_T(Z) / M_T(b\,3\ell)$ & $9.41\, 10^{-2}$ & $\Delta\varphi (t,\, Z)$ & $8.17\, 10^{-2}$ \\
$\eta^{\mbox{max}} (j)$ & $6.02\, 10^{-2}$ & $\Delta\varphi (\ell\ell|_Z)$ & $5.89\, 10^{-2}$\\
$\Delta\varphi (Z,\, \slashed{p}_T)$ & $5.37\, 10^{-2}$ & $p_T(j_1) / M_T(b\,3\ell)$ & $5.08\, 10^{-2}$\\
$\Delta\eta (\ell\ell|_Z)$ & $5.05\, 10^{-2}$ & $\Delta\eta (b,\, \ell_W)$ & $5.03\, 10^{-2}$ \\
$\eta(t)$ & $4.99\, 10^{-2}$ & $\Delta\varphi (Z,\, \ell_W)$ & $4.63\, 10^{-2}$\\
$\eta(Z)$ & $4.61\, 10^{-2}$ &  & \\ \hline
 \end{tabular}
 \caption{Ranking training variables for $M_{T'}=1.0$ TeV and full background. Here $\ell\ell|_Z$ identifies the 2 leptons that reconstruct the $Z$ boson.}
 \label{tab:MVA_var}
}

To define some of the angular variables, the whole four-momentum of the neutrino stemming from the semileptonic top-quark decay has been reconstructed as described above. Furthermore, we did not include the top mass (neither as invariant mass nor as transverse mass of its decay products) in the MVA because it did not show a strong discriminating power. This is understandable because signal and the largest sources of background both have a top quark in their intermediate states. Finally, the presence of a forward jet is a prominent feature of the signal. To account for this, we use the largest pseudorapidity of all jets $\eta^{\mbox{max}}(j)$ in the event.

It is interesting to notice that there are few variables which behaviour is directly proportional to the $T'$ mass, like the $p_T$ of the leading jet or the $p_T$ of the $2$ leptons reconstructing the $Z$ boson.
These correlations are efficiently removed if one consider ratios of the $p_T$'s over $M_T(b\,3\ell)$, which in fact decorrelate them. We checked that the highest sensitivity is reached when the ratio of said variables to the $T'$ reconstructed mass ($M_T(b\,3\ell)$) is considered instead of the actual observables. All other variables are almost uncorrelated, with a degree of correlation of $\pm30\%$ at most.

The variables in table~\ref{tab:MVA_var} are used to train the BDT to recognise the signal against the background. They are selected after the $Z$ mass reconstruction, i.e. after applying eq.~(\ref{eq:Z-cut}). A pictorial representation of these variables is in the Appendix, for the $M_{T'} = 1.0$ TeV with $R_L = 0.5$ benchmark point, and for the sum of all backgrounds. The BDT trained on each benchmark point is then applied on the full signal and background samples. A cut is performed on the BDT output to get the best signal-over-background significance. Values of the cuts and best significances are collected in table~\ref{tab:significances}.

We conclude this section with some further comments.
Firstly, we checked that the MVA does not suffer of overtraining as follows. Each sample (signal and background) is divided into 2 independent subsamples, one that is used for training and the other one for comparison. The absence of overtraining issues is shown in figure~\ref{fig:BDT_overtraining}: the output for the 2 subsamples coincides.

Secondly, we trained the MVA on the benchmark points (for $g^\ast = 0.1$ and $R_L=0.5$). In section~\ref{sect:results} we will make use of eq.~(\ref{eq:rescaling}) to extend the MVA analysis onto the whole $g^\ast$--$R_L$ plane, without the need to retrain the BDT any further. However, by controlling the share of the $T'$ coupling between first and third generation quarks, $R_L$ changes the ratio of incoming sea/valence quarks, which ultimately alters the process kinematics. We directly checked that the loss in performance, as compared to the results obtained with the suitable training, is at most of $\mathcal{O}(10\%)$ when $R_L$ goes to zero, thereby validating our extrapolating procedure.\footnote{Notice that the typical precision that one can aim at with a fast simulation is $\mathcal{O}(30\%)$.} The very same check is carried out for the cut-and-count analysis, with a similar performance behaviour.


Finally, the reader could wonder if the ratio of variables as described above might let the training procedure be less $M_{T'}$-dependent. It is in fact the case, but still it is not possible to use a universal MVA training and apply it to all the various $T'$ samples. This is because, as clear from table~\ref{tab:MVA_var}, still $M_T(b\,3\ell)$ is by far the best discriminating variable. 
If we remove it from the training, this becomes less $M_{T'}$-dependent for large $T'$ masses, but the overall performance of the MVA is even lower than previously. This let us conclude that it is not possible to create an efficient $M_{T'}$-independent training scheme.


\subsection{Results}\label{sect:results}
We collect here the final results for the discovery power at the LHC. 
In the case of the cut-and-count analysis of section~\ref{sect:cc}, we need to select a window around the signal peaks in the $M_T(b 3\ell)$ distribution. For the MVA of section~\ref{sect:MVA}, we need to perform a cut on the BDT output that maximises the significance. The maximum significance for the benchmark points, evaluated as $\sigma = S/\sqrt{S+B}$ after selecting a window around the mass peak or cutting on the BDT output, are collected in table~\ref{tab:significances}.

\TABLE{
\centering
\scalebox{0.84}{
 \begin{tabular}{|cc|c|c|c|c|c|}
 \hline
\multicolumn{2}{|c|}{Analysis} & $M_{T'}=0.8$ TeV & $M_{T'}=1.0$ TeV & $M_{T'}=1.2$ TeV & $M_{T'}=1.4$ TeV & $M_{T'}=1.6$ TeV\\
 \hline
\multicolumn{2}{|c|}{$M_T(b3\ell)$ cut (GeV)} & $[800-860]$    & $[840-1200]$    & $[1000-1340]$   & $[1120-1640]$ & $[1200-1800]$ \\
\multirow{3}{*}{C\&C} &S (ev.)  &18.00  &12.28  & 7.16  & 3.40  & 1.57 \\
		      &B (ev.)  & 8.90  & 4.88  & 1.74  & 0.90  & 0.63 \\
		      &$\sigma$ & 3.47  & 2.96  & 2.40  & 1.64  & 1.06 \\ \hline
\multirow{2}{*}{MVA} & cut   & 0.07 & 0.08 & 0.11 & 0.12 & 0.12 \\
 & $\sigma$                  & 3.64 & 3.10 & 2.50 & 1.62 & 1.15 \\  
 \hline
 \end{tabular}}
 \caption{Signal and background events and maximum significance for the benchmark points for $\mathcal{L}=100$ fb$^{-1}$, after selecting a mass window (for the C\&C), or after cutting on the BDT output (MVA).}
 \label{tab:significances}
}

One of the most important result in this paper is that the dedicated BDT analysis does not significantly improve on the cut-and-count strategy, as clear from table~\ref{tab:significances}. The latter analysis is certainly sufficient and easier. The cuts as displayed in eqs.~(\ref{eq:W-cut})--(\ref{eq:top-cut}) are already best optimised, as is the signal peak selection. No further variable/cut need to be considered/applied.

The significances in table~\ref{tab:significances} are for the benchmark points, evaluated for $g^\ast = 0.1$ and $R_L=0.5$. We can now extrapolate them to the full $g^\ast$--$R_L$ parameter space using eq.~(\ref{eq:rescaling}). The extrapolation is done by rescaling the cross section and every time reoptimising the cuts to get the highest significance. The $3$ and $5$ sigma discovery lines are drawn as a function of $g^\ast$ and the $T'$ mass for some fixed values of $R_L$ in figure~\ref{fig:significance_100}(left), and as a function of $g^\ast$ and $R_L$ for the benchmark $T'$ masses in figure~\ref{fig:significance_100}(right).
figure~\ref{fig:significance_100} shows that with $100$ fb$^{-1}$ of data, $T'$ masses up to $2$ TeV can be observed, depending on the values of the couplings. The cross section for the trilepton decay channel of the $T'$ (and hence the LHC reach) increases considerably when $R_L$ is non-vanishing, getting to a maximum for $R_L\simeq 1$, corresponding to $50\%$--$50\%$ mixing. This effect is simply due to the increased admixture of valence quarks in production, mitigated by a reduced $T'$-to-$tZ$ branching ratio, as $R_L$ increases.

\FIGURE{
\centering
\includegraphics[width=0.48\linewidth]{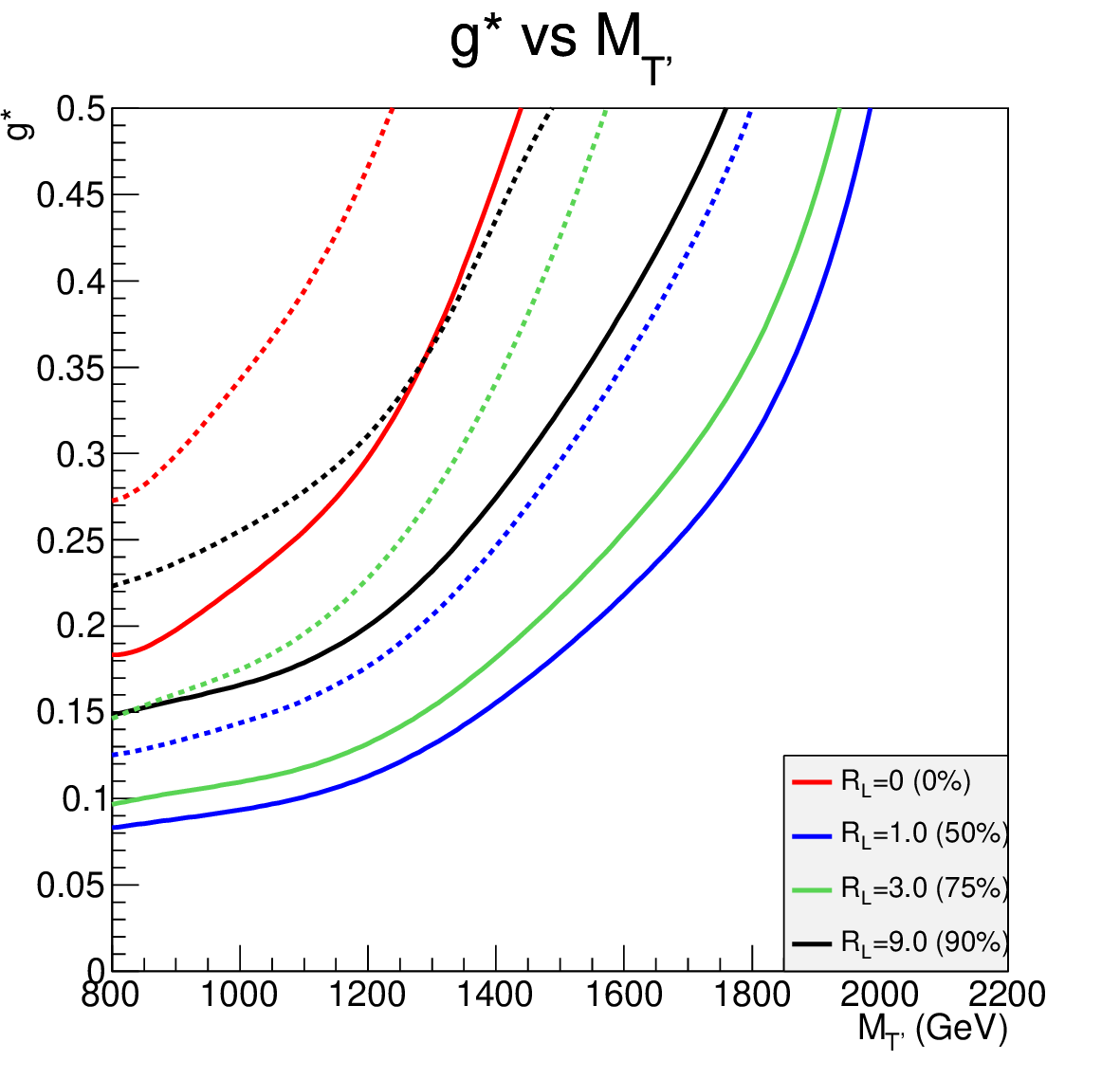} 
\includegraphics[width=0.48\linewidth]{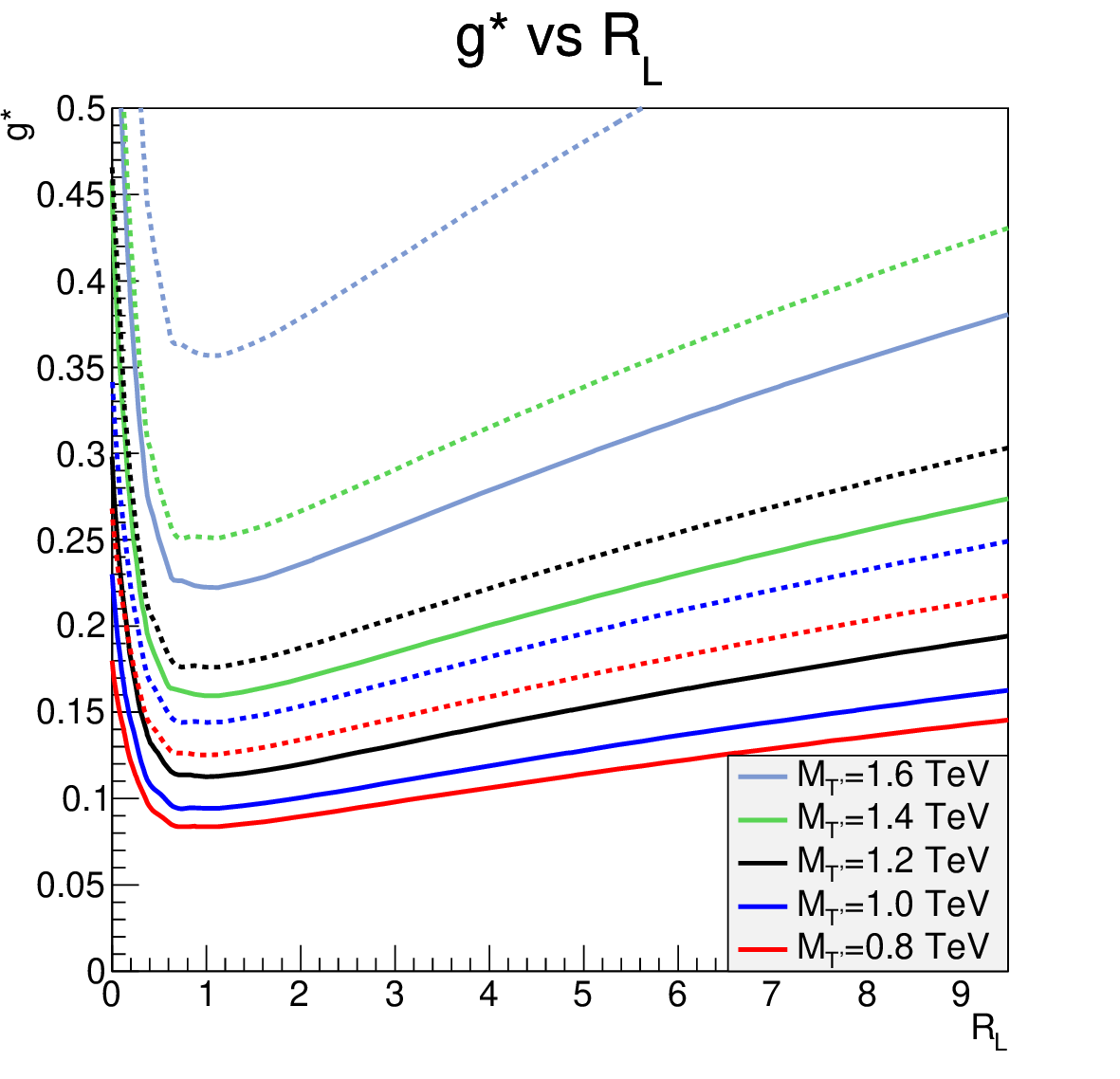} 
\caption{Significance $\sigma=3$ (solid lines) and $\sigma=5$ (dashed lines) for $\mathcal{L}=100$ fb$^{-1}$.}
\label{fig:significance_100}
}

The reach in $g^\ast$ is here roughly twice than for the no mixing case ($R_L=0$).
Then, for larger values of $R_L$, $g^\ast$ needs to slightly increase to compensate for the decrease in cross section due to the larger mixing with the first generation quarks, that suppress the $T'\to tZ$ branching ratio.

The discovery power of the trilepton channel can be compared to the one of the dilepton channel as studied in~\cite{Reuter:2014iya}. The $R_L=0$ line in figure~\ref{fig:significance_100} is the one considered therein. However, to be able to draw a meaningful comparison, we shall set ourselves in the same conditions,\footnote{ref.~\cite{Reuter:2014iya} used an integrated luminosity of $300$~fb$^{-1}$ and rescaled the signal by a mean $k$-factor of $1.14$.} which correspond to the end of the LHC run-II. The plot for this setup is in figure~\ref{fig:significance_300}. The curve to be compared is the $R_L=0$ one on the left-hand side plot. At low $T'$ masses, the dileptonic channel of ref.~\cite{Reuter:2014iya} performs slightly better, meaning that a marginally lower value of $g^\ast$ can be probed. At larger $T'$ masses though the trileptonic channel is more sensitive, extending the reach by $200-300$ GeV. In these conditions, our analysis is sensitive to $g^\ast$ couplings down to $0.05$ and $T'$ masses up to $2.1$ TeV at most.
\FIGURE{
\centering
\includegraphics[width=0.48\linewidth]{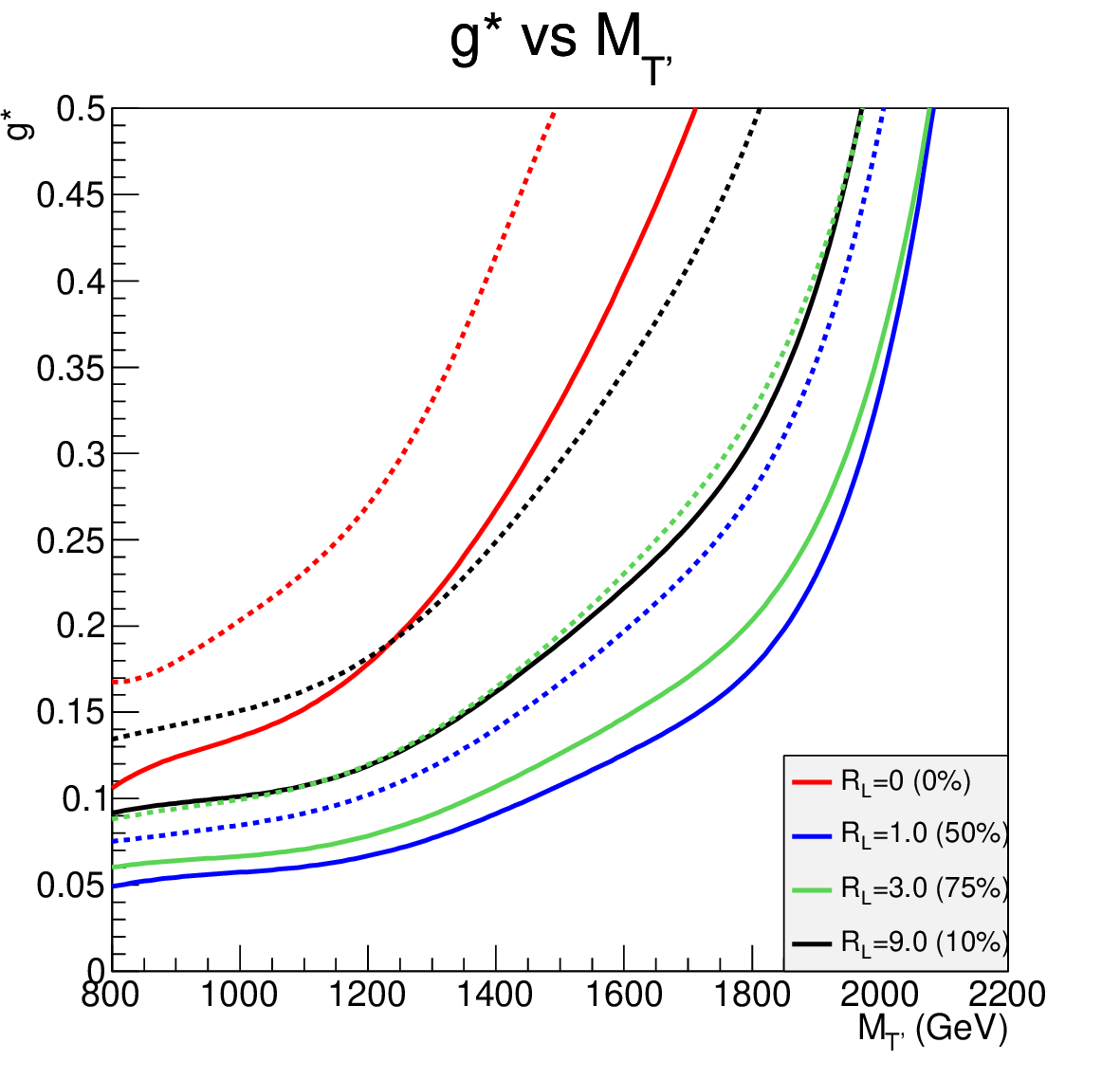} 
\includegraphics[width=0.48\linewidth]{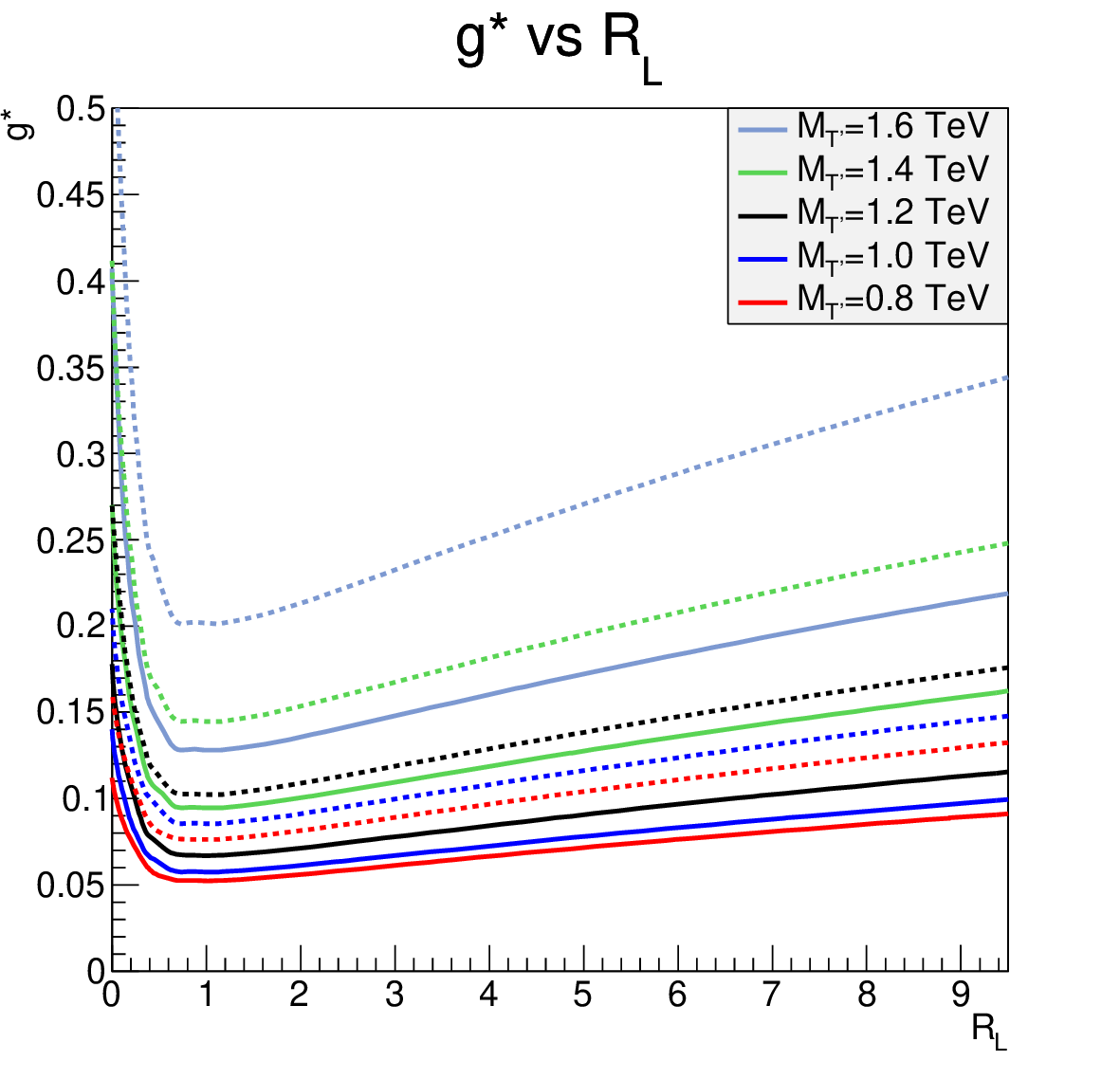} 
\caption{Significance $\sigma=3$ (solid lines) and $\sigma=5$ (dashed lines) for $\mathcal{L}=300$ fb$^{-1}$ and $k_f=1.14$ as in ref.~\cite{Reuter:2014iya}.}
\label{fig:significance_300}
}

\subsection{Top FCNC reinterpretation}\label{sect:FCNC}
We conclude this paper by presenting a reinterpretation of our investigation in terms of the top-quark FCNC coupling to a light quark and a $Z$ boson. In this scenario, the top quark interacts with a $Z$ boson and a up- or charm-quark via the $\kappa_{tZq}$ FCNC coupling~\cite{AguilarSaavedra:2008zc}
\begin{equation}\label{eq:kzqt}
\mathcal{L}= \sum_{q=u,c} \frac{g}{\sqrt{2}c_W} \frac{\kappa_{tZq}}{\Lambda}\,\overline{t}\sigma^{\mu\nu}\,\big(f^L_{Zq} P_L + f^R_{Zq} P_R \big)\,q\, Z_{\mu\nu}\, ,
\end{equation}
where $\Lambda$ is the scale of new physics.
The Lagrangian in eq.~(\ref{eq:kzqt}) gives a similar final state as the one subject of this paper, $pp\to tZ$, with a top-quark and a $Z$ boson produced back-to-back.
The only difference with the $T'$-induced topology is the absence of the forward jet at leading order. The analyses of the $T'$-mediated signature subject of this paper could therefore be as well sensitive to the one induced by the top effective coupling. We tested it by producing at leading order a $pp\to tZ$ sample when turning on one FCNC coupling at the time. We label them $\kappa_{tZu}$ and $\kappa_{tZc}$, respectively. The samples have been produced as described in section~\ref{sect:pheno}. Finally, they have been analysed at detector level by running them on the cut-and-count analysis of section~\ref{sect:cc}. 
Efficiencies and event yields for $100$~fb$^{-1}$ are collected in table~\ref{tab:fcnc_eff}.

\begin{table}[!t]
\centering
 \begin{tabular}{|c|c|c|}
 \hline
 Cut  & $\kappa_{tZu}$  & $\kappa_{tZc}$ \\
 \hline
 no cuts &  $2263(100\%)$ & $5360(100\%)$\\
 $1\leq n_j\leq 3$ & 1765(78.0\%) & 4452(83.0\%)  \\
 $n_\ell \equiv 3$ & 191.8(10.9\%) & 623.3(14.0\%) \\
 $n_b \equiv 1$ & 113.8(59.3\%)	 & 381.0(61.1\%) \\
 cut~(\ref{eq:Z-cut})   & 103.2(90.7\%)  & 342.7(90.0\%) \\
 cut~(\ref{eq:W-cut})   & 96.2(93.3\%)   & 323.6(94.4\%)\\
 cut~(\ref{eq:top-cut}) & 91.1(94.7)     & 304.7(94.1\%)\\ \hline
$M_T(b3\ell)$ & $>400$ GeV & $>200$ GeV \\ \hline
 S & 68.0  & 304.5\\
 B & 102.9 & 241.7\\
 $\sigma$ & 5.2 & 13.0\\ \hline
 \end{tabular}
 \caption{Surviving events (and efficiencies with respect to the previous item) for the cut-and-count analysis of the FCNC top coupling $\kappa_{tZu}/\Lambda$ (at current limit) and $\kappa_{tZc}/\Lambda$ (for BR$(t\to Zc)=1\%$), and signal/background events that maximise the significance.}
 \label{tab:fcnc_eff}
\end{table}


The $M_T(b 3\ell)$ distribution after the application of the cuts of eqs.~(\ref{eq:Z-cut})--(\ref{eq:top-cut}) is shown in figure~\ref{fig:MtZ_fcnc}. The significance for the $\kappa_{tZu}$ sample is maximised by selecting $M_T(b 3\ell) > 400$ GeV, reaching the value of $5.2$ sigma for the present best limit of the coupling of $\kappa_{tZu}/\Lambda= 0.2$~TeV$^{-1}$ (or BR$(t\to Zu)= 0.05\%$)~\cite{Chatrchyan:2013nwa}, corresponding to a cutoff scale $\Lambda=5$ TeV. Regarding the $\kappa_{tZc}$ sample, we chose a coupling yielding BR($t\to Zc)=1\%$ to compare the results. For this value, the highest significance of $13.0\sigma$ is obtained by selecting $M_T(b 3\ell) > 200$ GeV. Notice that the only available limit to this coupling is such that BR$(t\to Zc)\leq 11.4\%$~\cite{CMS-PAS-TOP-12-021}. For comparison, we applied the MVA trained on each $T'$ signal to the FCNC case. Also in this case however it did not improve the sensitivity.

\FIGURE{
\centering
\includegraphics[width=\linewidth]{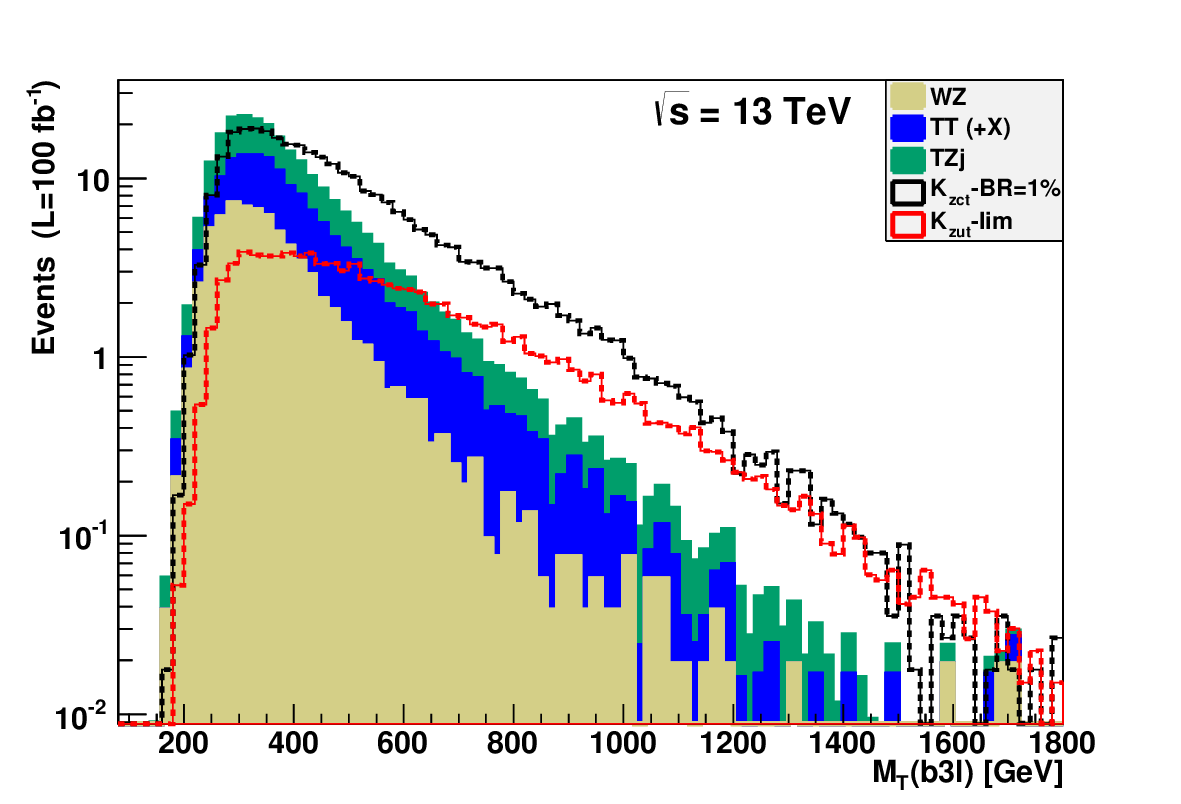} 
\caption{$M_T(b\,3\ell)$ distribution with the present best limit on the top-Z-up FCNC coupling and for BR$(t\to Zc)=1\%$ for the top-Z-c one.}
\label{fig:MtZ_fcnc}
}

Finally, the cut-and-count discovery power curves at the LHC as a function of the top FCNC couplings for $100$~fb$^{-1}$ are shown in figure~\ref{fig:sig_fcnc}. Contours for a significance $\sigma=2,3,5$ are presented. This analysis can gather evidences for the existence of FCNC processes in the top sector for $\kappa_{tZu}/\Lambda$($\kappa_{tZc}/\Lambda$) couplings leading to BRs above $0.025\%(0.16\%)$ for $100$~fb$^{-1}$. In case of non observation, $95\%$ C.L. limits can be put for BRs down to $0.016\%$ and $0.1\%$ for the two couplings, respectively.

\FIGURE{
\centering
\includegraphics[width=0.75\linewidth]{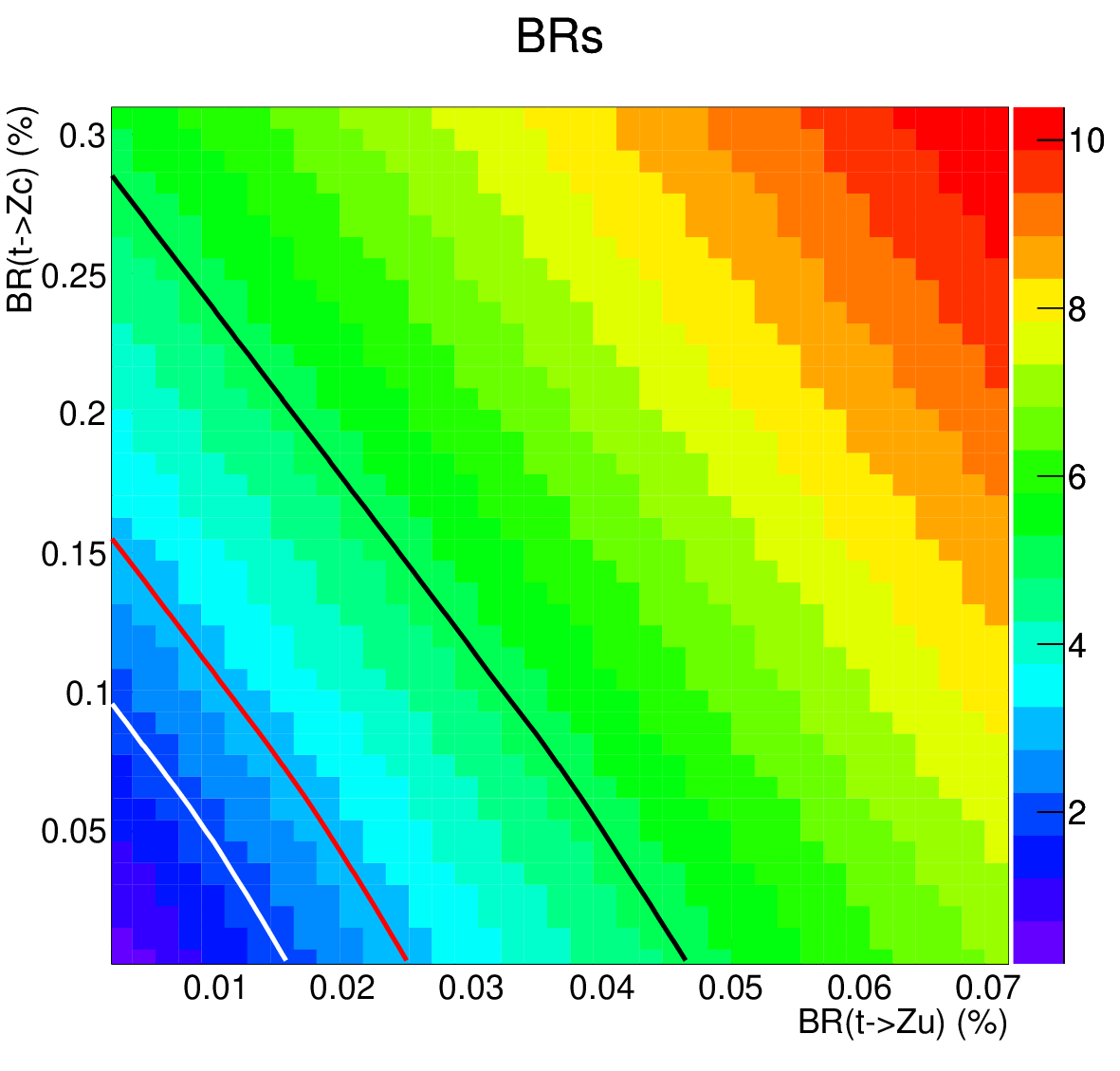} 
\caption{Significance as a function of the top FCNC couplings for the cut-and-count analysis, and $\sigma=2$(white), $\sigma=3$(red), and $\sigma=5$(black) contour lines.}
\label{fig:sig_fcnc}
}

\section{Conclusions}\label{sect:conclusions}
In this work we described the LHC run-II discovery potential of the trilepton channel for a singlet top partner in the single production mode and its subsequent decay into a top quark and a $Z$ boson. A simple cut-and-count analysis has been designed, by selecting and cutting the most straightforward distributions. A suitable multivariate analysis did not improve significantly on the cut-and-count results. The comparison was performed on several signal benchmark points. Further, we proposed a simple way to extend our results to the whole parameter space of a simplified model. 

Overall, a search at the LHC in the trilepton channel can be sensitive to top partners decaying into $tZ$ for masses up to $2.0(2.1)$ TeV and couplings down to $0.08(0.05)$ with $100(300)$ fb$^{-1}$ of data. We compared to the reach in the dilepton channel of ref.~\cite{Reuter:2014iya}, concluding that the trilepton mode can extend the former by $200-300$ GeV in $T'$ masses for suitable values of the couplings. Finally, we reinterpreted our analyses in the context of a top FCNC coupling to a $Z$ boson and a light quark, which provides a similar final state. We showed that this channel can discover at $5\sigma$ values of the couplings at the present best exclusion limit (for $100$~fb$^{-1}$), probe at $3\sigma$ FCNC branching ratios down to $0.025\%(0.16\%)$ for $\kappa_{tZu}/\Lambda$($\kappa_{tZc}/\Lambda$), or eventually extend the exclusion limits down to 
$0.016\%$ and $0.1\%$ for the two FCNC couplings, respectively.

\section*{Acknowledgements}
We would like to sincerely thank our colleagues A.~Alloul, C.~Collard, E.~Conte and G.~Hammad for the help in generating the background samples used in this work and for useful comments. The work of LB is supported by the Theorie-LHC France initiative of the CNRS/IN2P3 and by the French ANR 12 JS05 002 01 BATS@LHC.

\newpage

\appendix
\section{BDT variables}

We present here the training variables for the $M_{T'} = 1.0$ TeV with $R_L = 0.5$ benchmark point. The overtraining test is shown in figure~\ref{fig:BDT_overtraining}. As explained in the text, it shows that there is no overtraining issue.

\begin{figure}[!ht]
\centering
\includegraphics[width=0.3\linewidth]{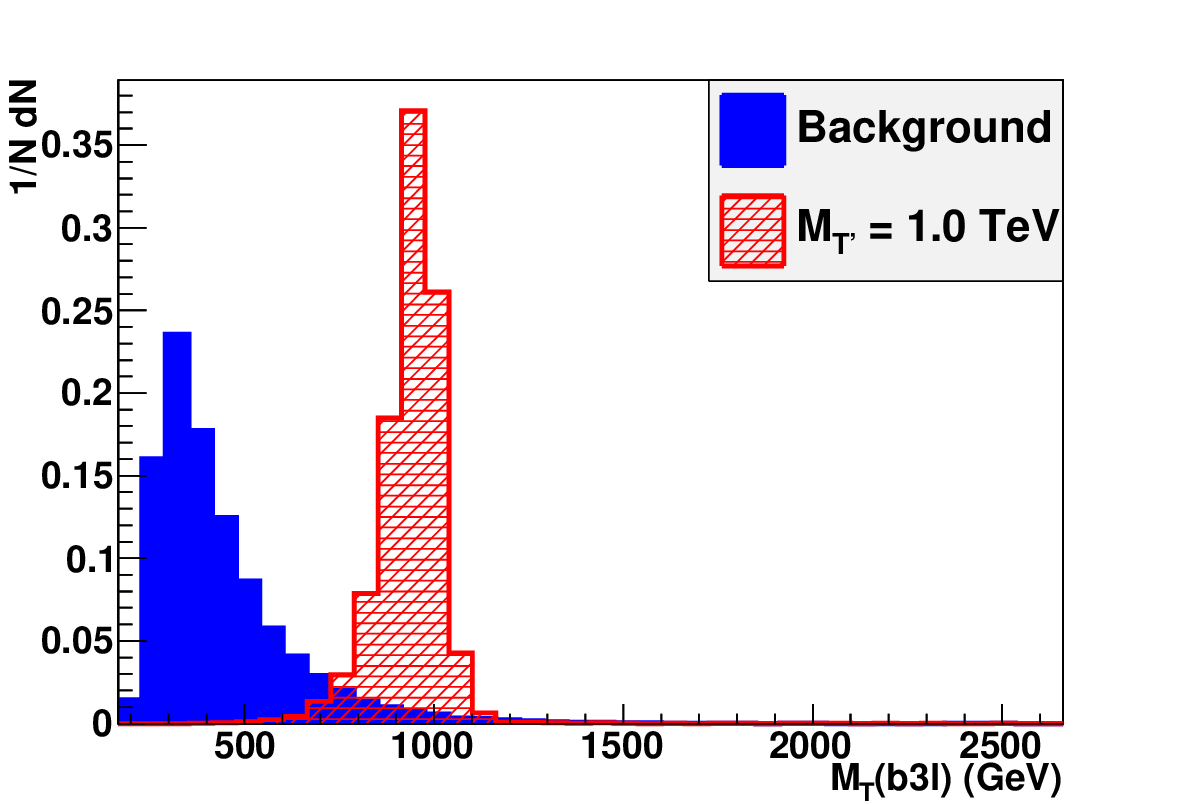} 
\includegraphics[width=0.3\linewidth]{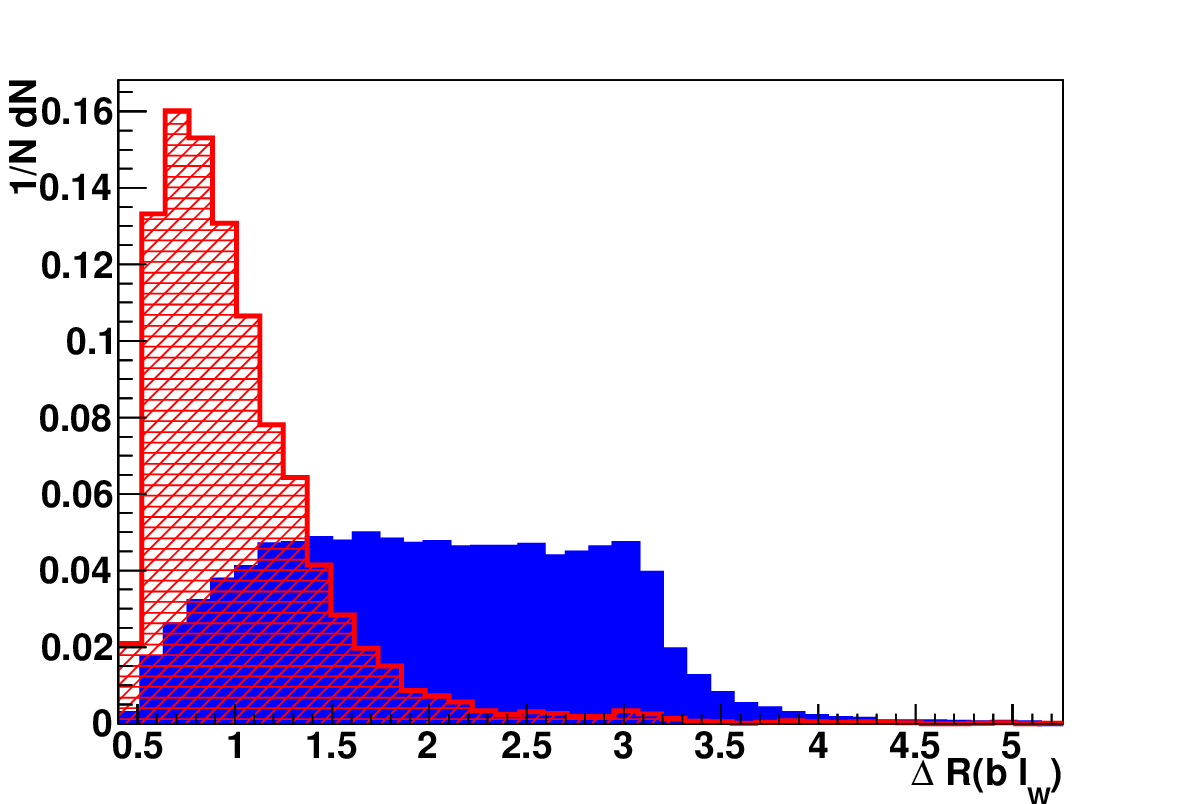} 
\includegraphics[width=0.3\linewidth]{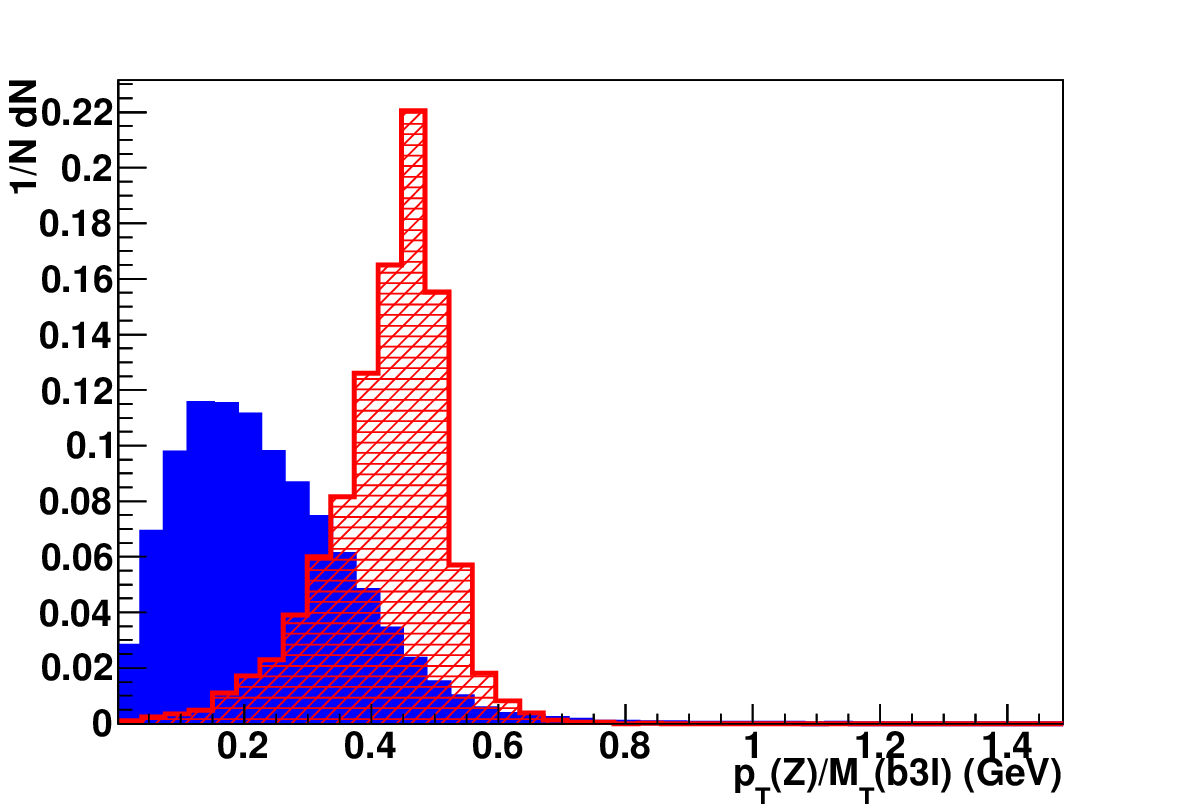} \\
\includegraphics[width=0.3\linewidth]{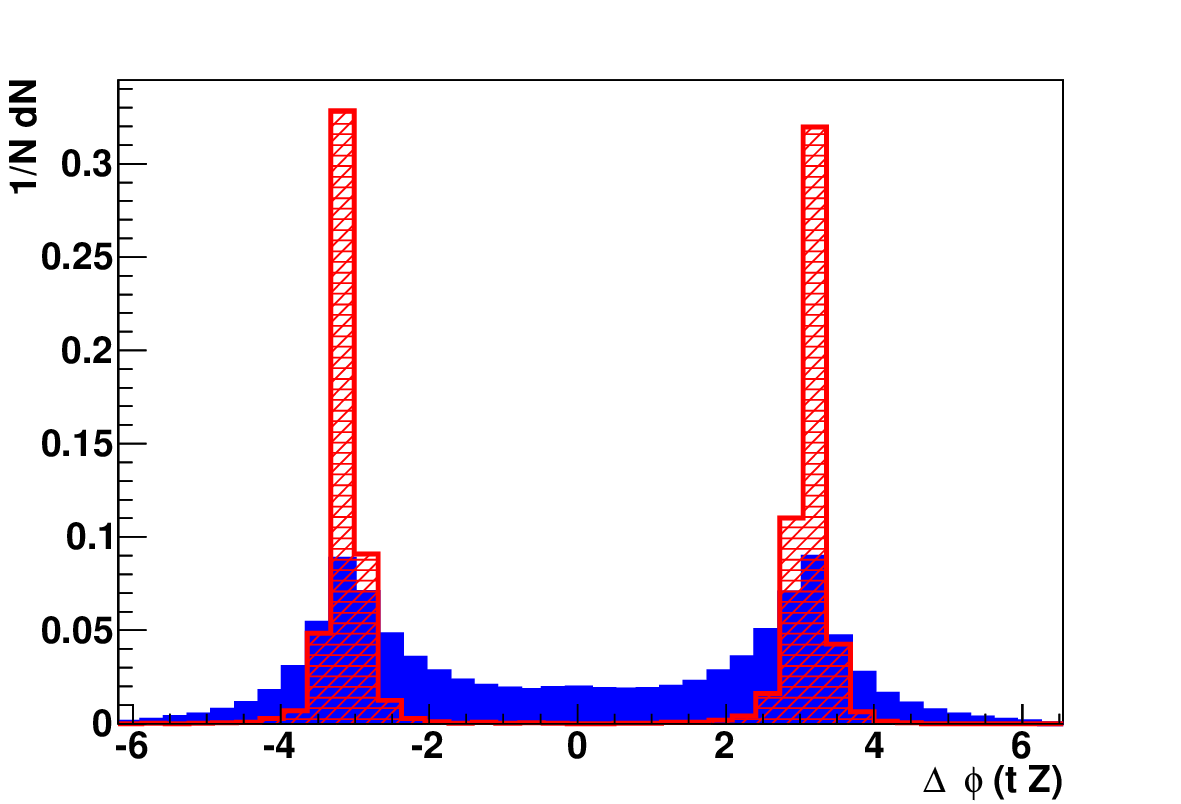} 
\includegraphics[width=0.3\linewidth]{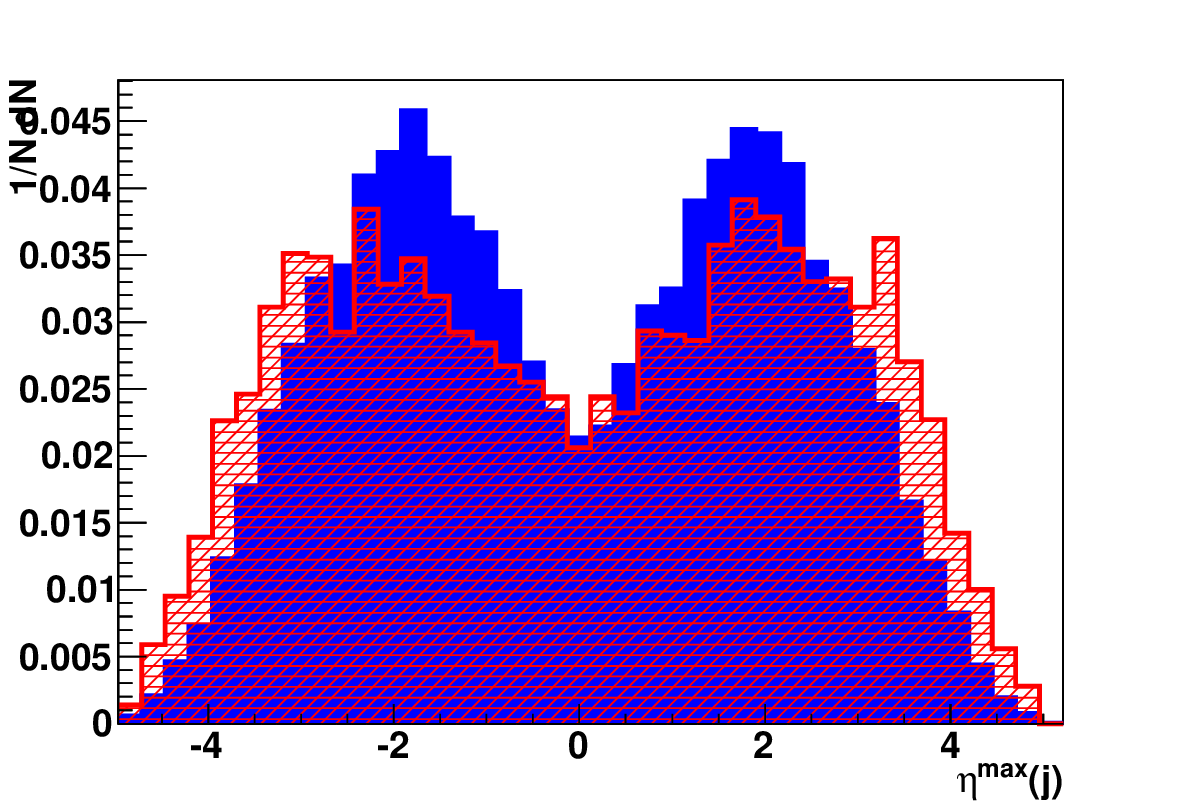} 
\includegraphics[width=0.3\linewidth]{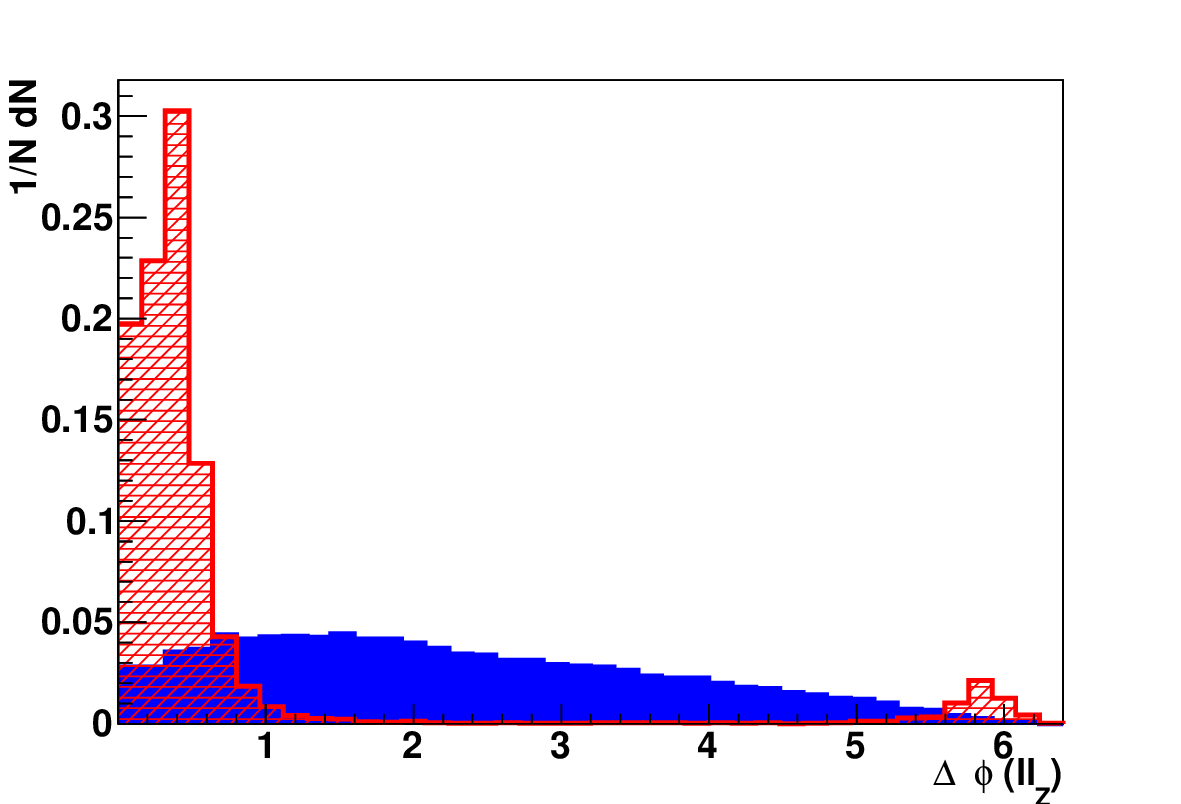}
\caption{BDT training variables (1).}
\label{fig:BDT_training_1}
\end{figure}

\begin{figure}[!ht]
\centering
\includegraphics[width=0.3\linewidth]{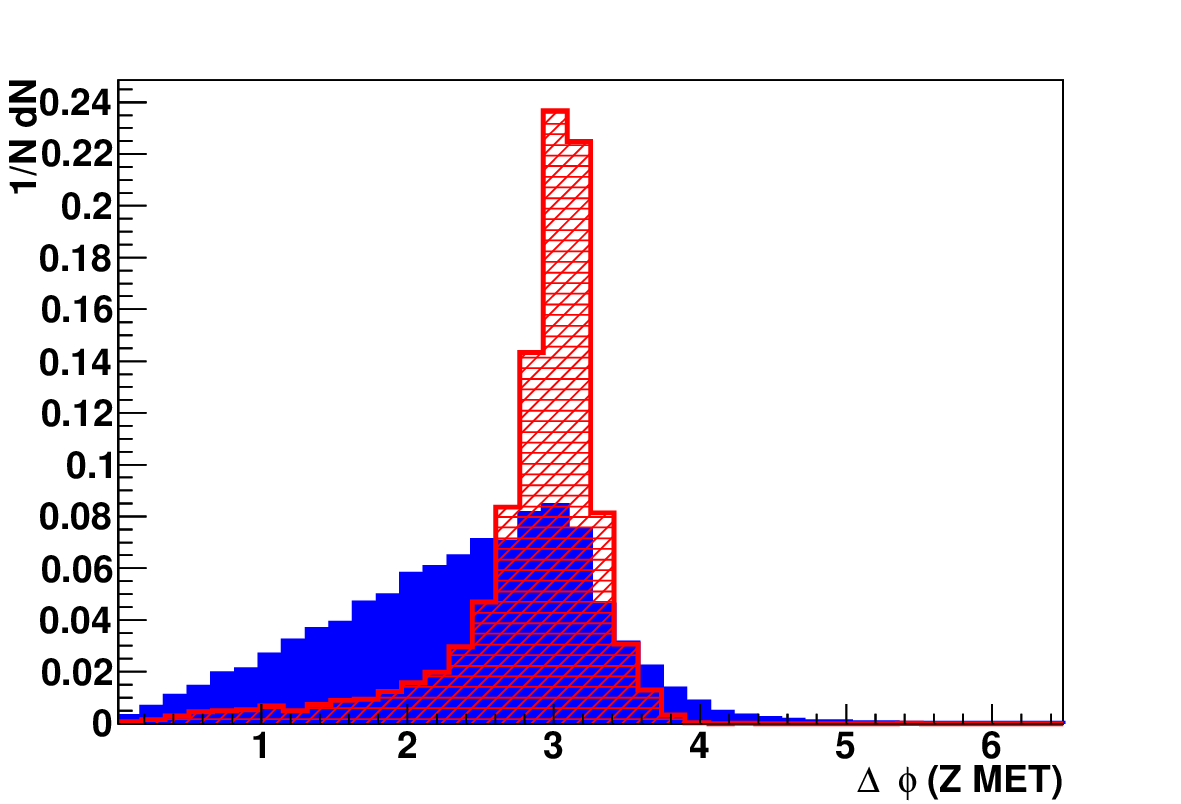} 
\includegraphics[width=0.3\linewidth]{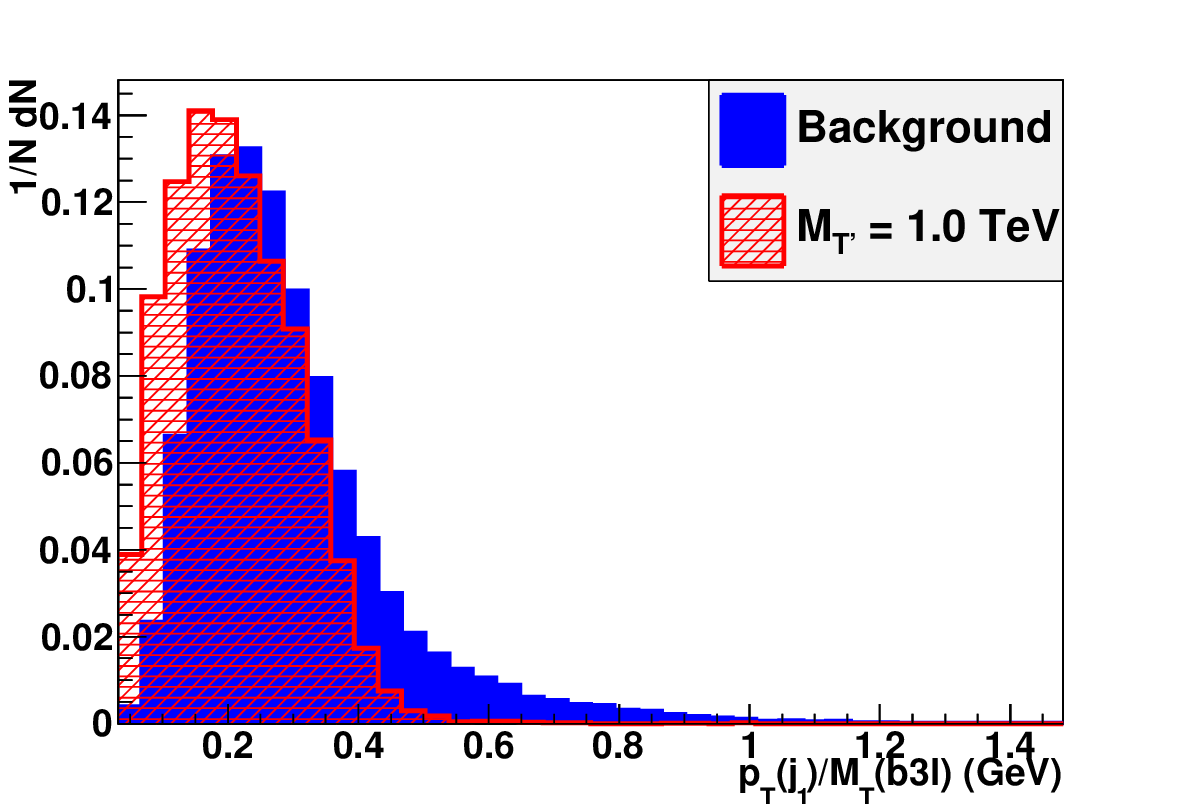} 
\includegraphics[width=0.3\linewidth]{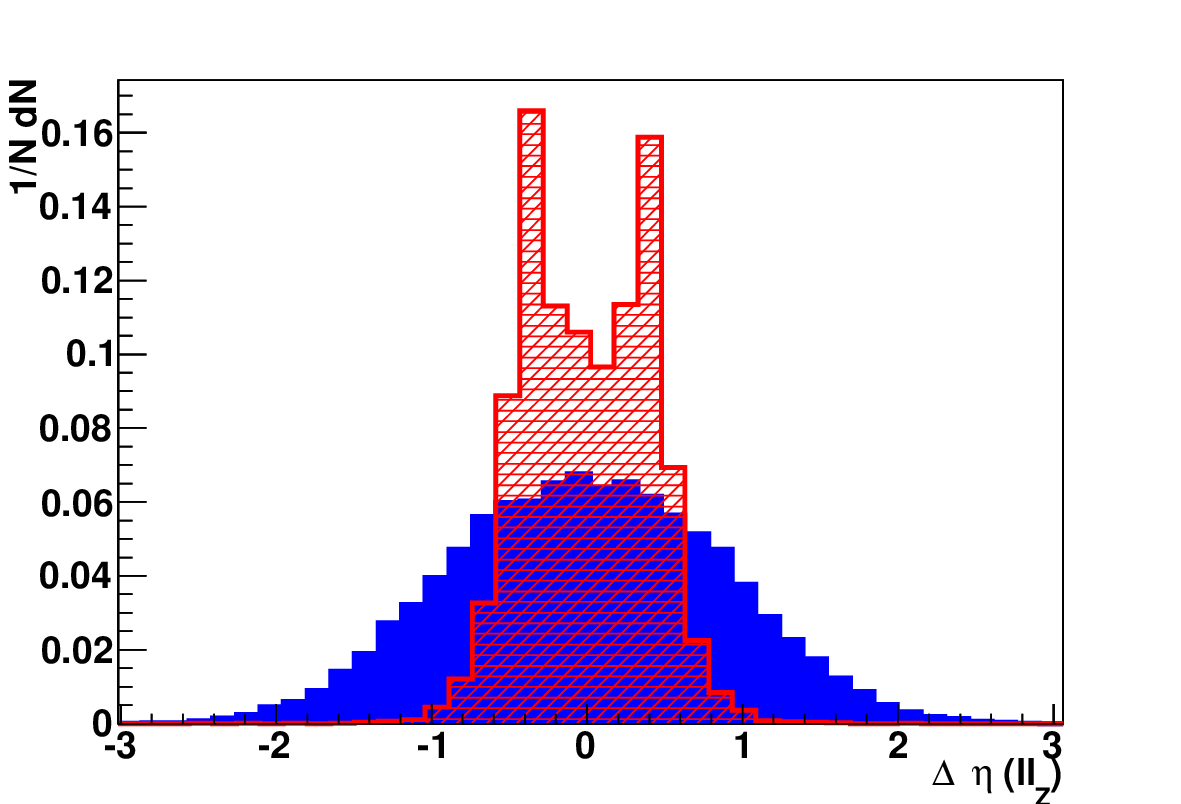} \\
\includegraphics[width=0.3\linewidth]{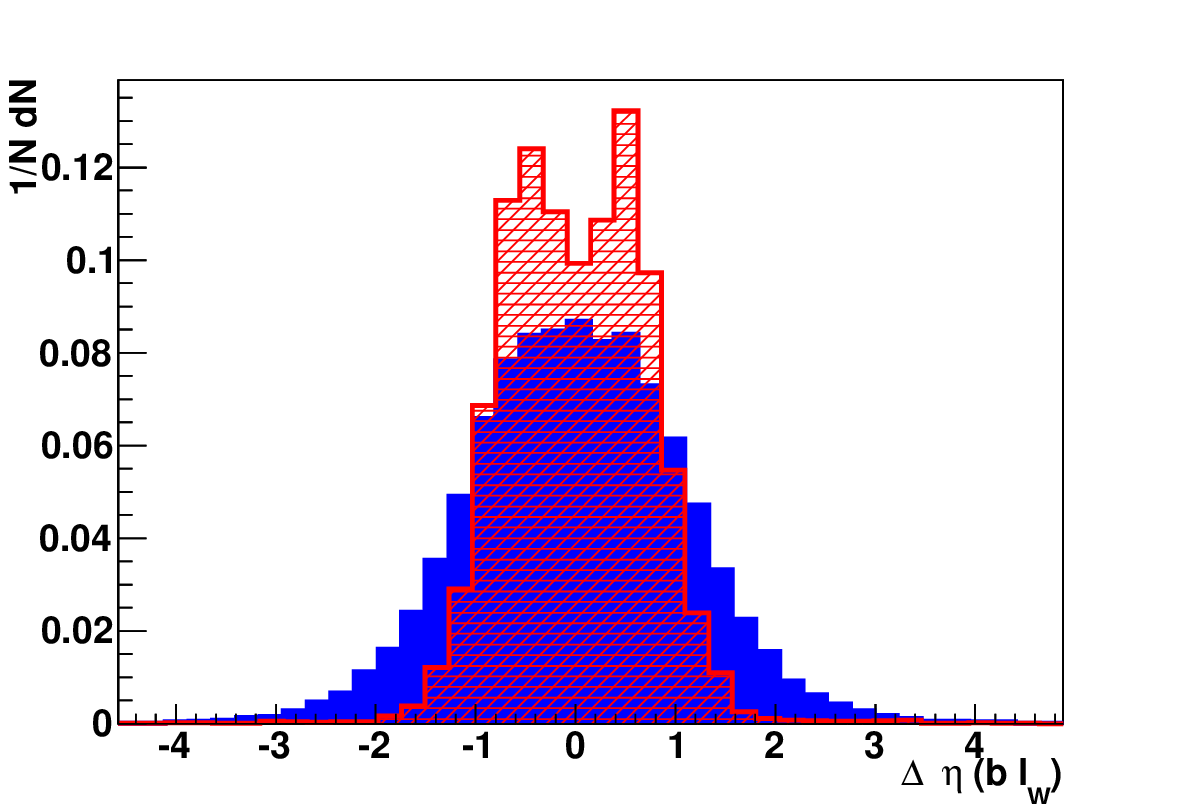} 
\includegraphics[width=0.3\linewidth]{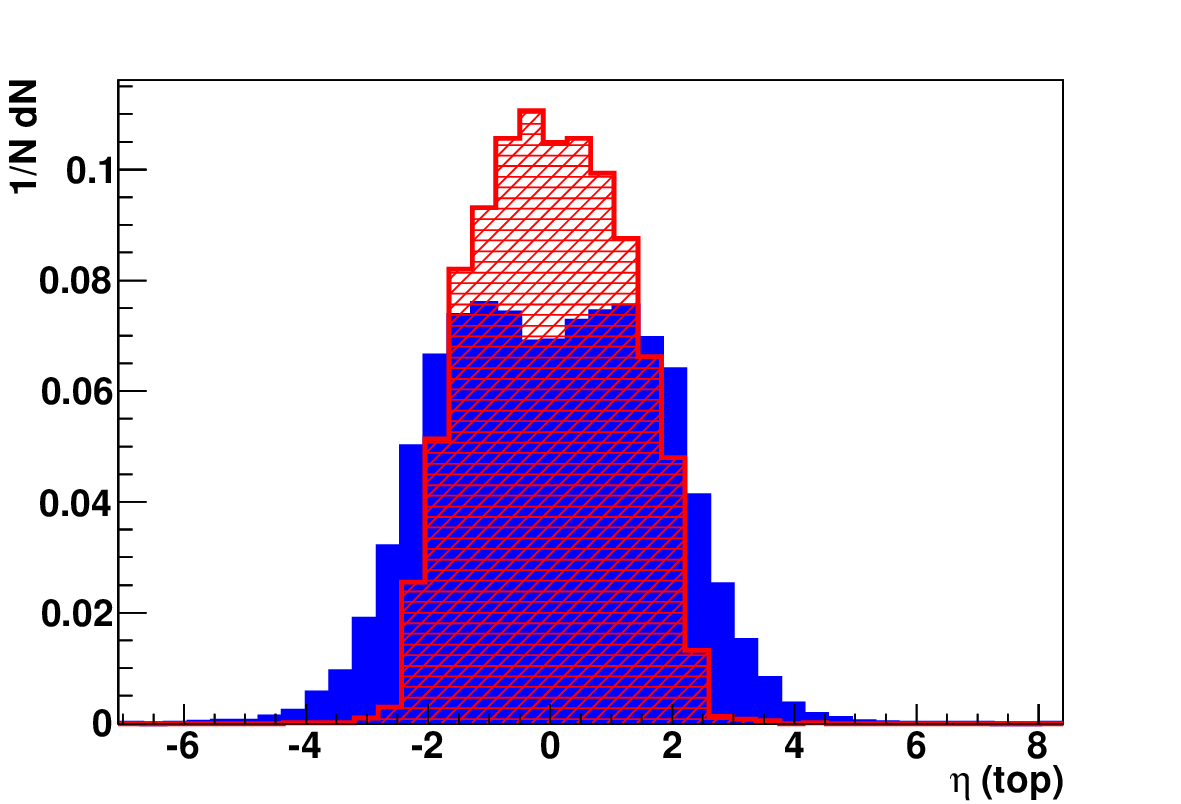} 
\includegraphics[width=0.3\linewidth]{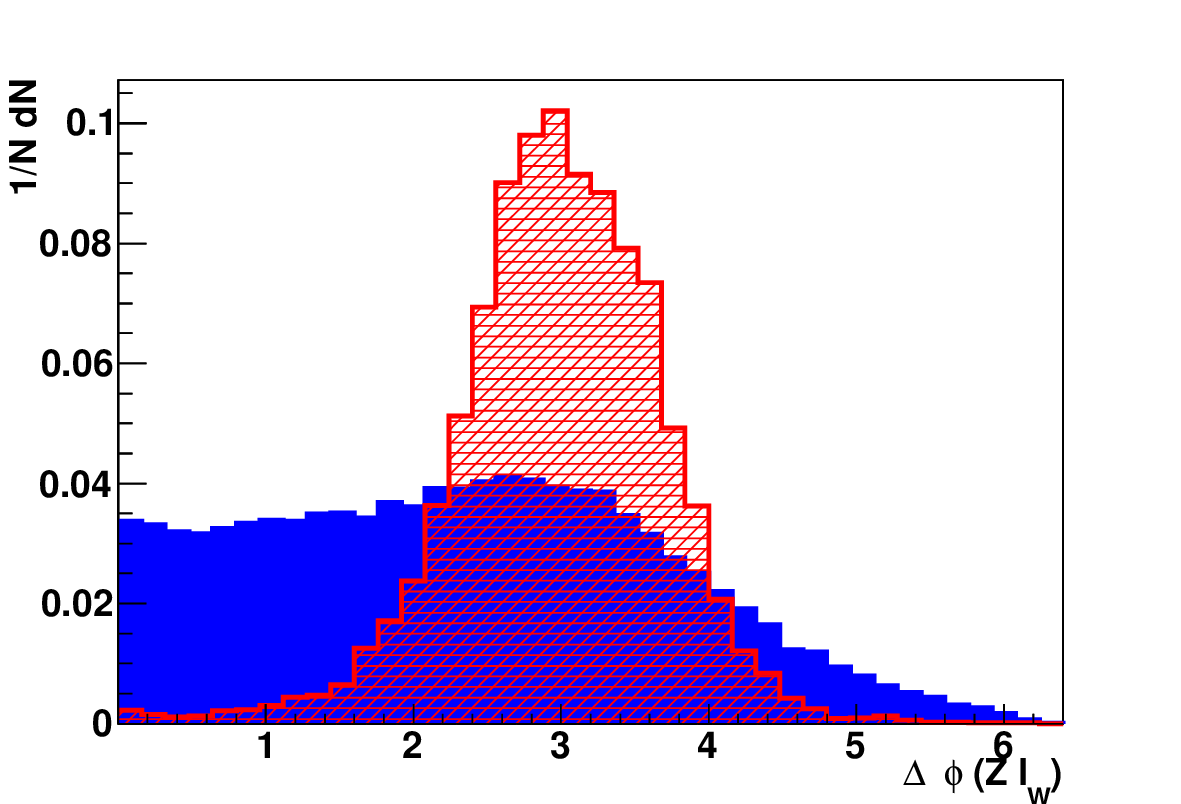} \\
\includegraphics[width=0.3\linewidth]{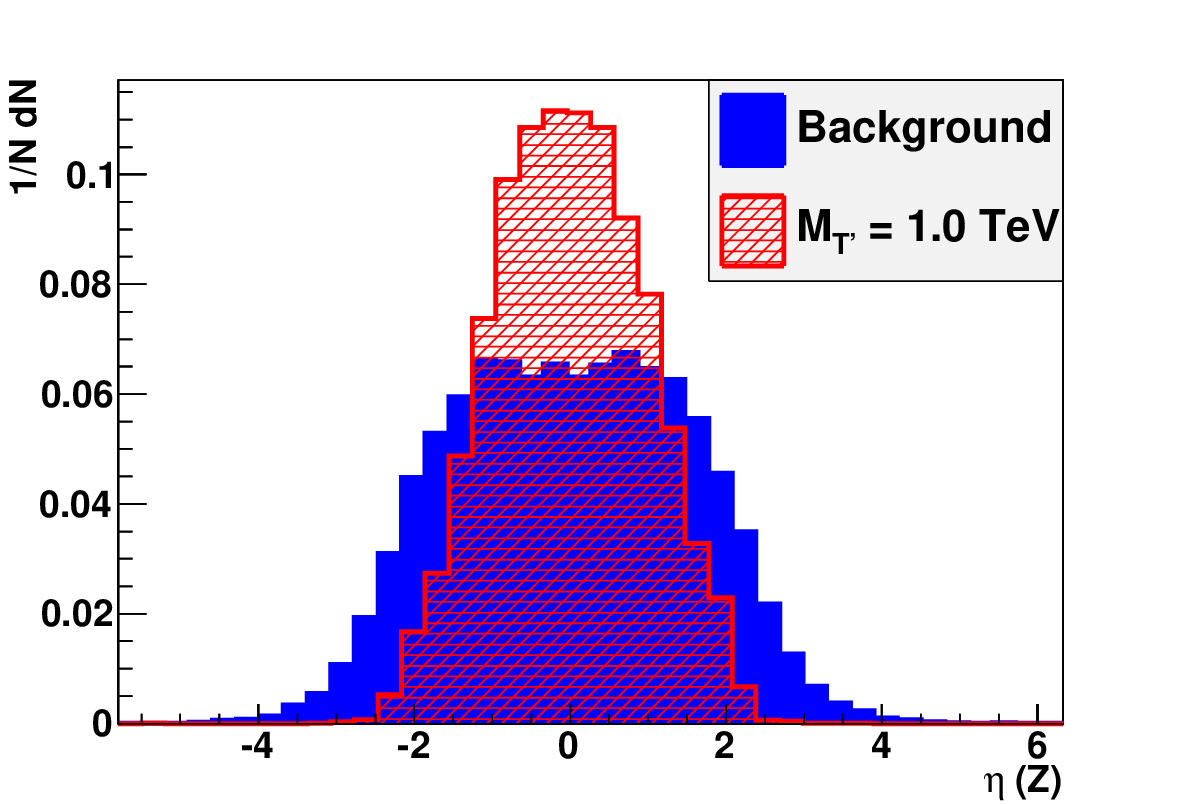} 
\caption{BDT training variables (2).}
\label{fig:BDT_training_2}
\end{figure}

\FIGURE{
\centering
\includegraphics[width=0.8\linewidth]{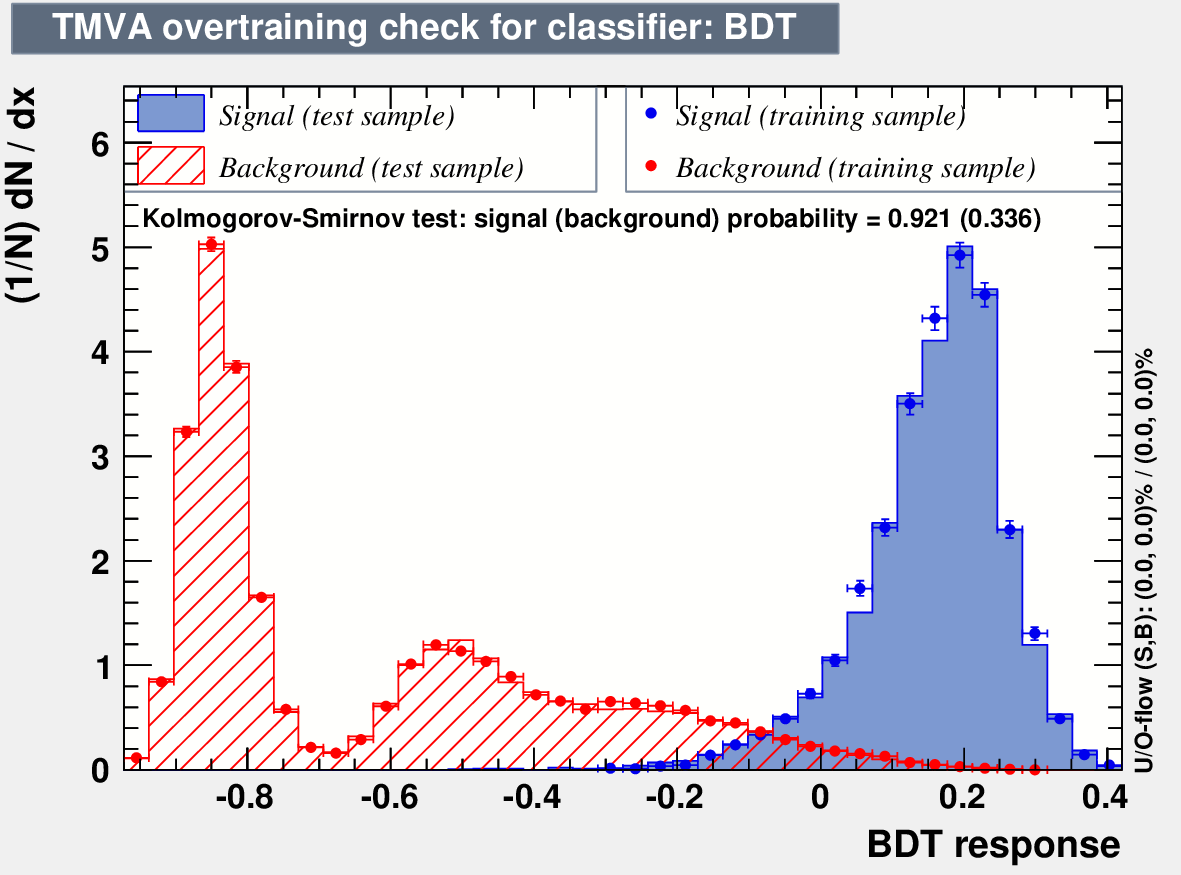} 
\caption{BDT overtraining.}
\label{fig:BDT_overtraining}
}

\bibliographystyle{h-physrev5}
\bibliography{Tp3l}

\end{document}